\documentclass[fp,twocolumn,dvipdfmx]{jpsj3}

\usepackage{graphicx,color}

\usepackage{amsmath,amssymb,nccmath,rmath2}
\usepackage{widetext}
\usepackage{subfigure}
\usepackage[utf8]{inputenc}
\usepackage[T1]{fontenc}
\usepackage{ulem}
\usepackage[abs]{overpic}

\renewcommand{\eqref}[1]{Eq. (\ref{#1})}
\newcommand{\figref}[1]{Fig. \ref{#1}}

\newcommand{\refcite}[1]{Ref. \citen{#1}}
\newcommand{\appnref}[1]{Appendix \ref{#1}}

\newcommand{\seqeqref}[2]{Eqs. (\ref{#1}) and (\ref{#2})}
\newcommand{\seqfigref}[2]{Figs. \ref{#1} and \ref{#2}}

\newcommand{\eqrefInit}[1]{Equation (\ref{#1})}
\newcommand{\figrefInit}[1]{Figure \ref{#1}}

\newcommand{\etal}{\textit{et al.}}
\newcommand{\ie}{\textit{i.e.}}
\newcommand{\eg}{\textit{e.g.}}


\title{Analysis of the Hopfield model with Discrete Coupling}
\author{
	Ryuta Sasaki$^1$\thanks{sasaki.r.al@m.titech.ac.jp}, and
	Toru Aonishi$^1$\thanks{aonishi@acs.dis.titech.ac.jp}
}
\inst{$^1$ School of Computing, Tokyo Institute of Technology, Tokyo, Japan}

\abst{
	Growing demand for high-speed Ising-computing-specific hardware
has prompted a need for determining how the accuracy depends on a hardware implementation with physically limited resources. 
For instance, in digital hardware such as field-programmable gate arrays, as the number of bits representing the coupling strength is reduced, 
the density of integrated Ising spins and the speed of computing can be increased while the calculation accuracy becomes lower. 
To optimize the accuracy-efficiency trade-off, we have to estimate the change in performance of the Ising computing machine depending on the number of bits representing the coupling strength. 
In this study, we tackle this issue by focusing on the Hopfield model with discrete coupling.
The Hopfield model is a canonical Ising computing model. 
Previous studies have analyzed the effect of a few nonlinear functions (\eg~sign) for mapping the coupling strength on the Hopfield model with statistical mechanics methods, 
but not the effect of discretization of the coupling strength in detail. 
Here, we derived the order parameter equations of the Hopfield model with discrete coupling by using the replica method
and clarified the relationship between the number of bits representing the coupling strength and the critical memory capacity.
In this paper,
we used the replica method for the Hopfield model with general nonlinear coupling (Sompolinsky (1986))
to analyze the model with a multi-bit discrete coupling strength,
and we novelly derived the de Almeida-Thouless line
of the model with general nonlinear coupling.

}

\begin{document}

\maketitle
\thispagestyle{jpsj}

\section{Introduction}

Combinatorial optimization problems are ubiquitous in many fields, 
such as 
traffic optimization, \cite{neukart2017traffic}
scheduling and planning, \cite{rieffel2015case,venturelli2015quantum}
resource allocation, \cite{kwak2018simulated}
drug design, \cite{kitchen2004docking,sakaguchi2016boltzmann} 
portfolio optimization, \cite{rosenberg2016solving}
and machine learning. \cite{crawford2016reinforcement,khoshaman2018quantum,henderson2018leveraging,levit2017free}
Many important problems
belong to the nondeterministic polynomial time (NP)-hard complexity class,
and for typical instances, require a computation time that scales exponentially with the problem size.
Many of these problems can be translated into problems of finding the ground states of an Ising model. \cite{lucas2014ising}
The Hamiltonian of an Ising model is written as
\begin{align}
	\ham (S_1, \dots, S_N)
	= - \cfrac{1}{2} \sum_{i \neq j} J_{ij} S_i S_j,
	\label{eq:hamiltonian}
\end{align}
where $S_1,\dots,S_N$ are Ising variables, which take either $-1$ or $+1$,
and $J_{ij}$ expresses the coupling strength between the $i$th and $j$th Ising variables.
The coupling strength is symmetric, \ie, $J_{ij} = J_{ji}$.

The mapping from many combinatorial optimizations onto an Ising model
motivated us to develop machines dedicated to the search for the ground state.
Many such machines have been proposed in the past decade.
Well-known examples are the hardware devices of D-Wave Systems Inc. \cite{johnson2011quantum}
These devices use quantum annealing, \cite{kadowaki1998quantum,brooke1999quantum,santoro2002theory,das2008colloquium}
or quantum adiabatic computation. \cite{farhi2001quantum,albash2018adiabatic}
Utsunomiya \etal~proposed a coherent Ising machine (CIM) that executes Ising computations in an injection-locked laser network. \cite{utsunomiya2011mapping}
The CIM is now based on degenerate parametric oscillators. \cite{
	wang2013coherent,
	marandi2014network,
	inagaki2016coherent,
	mcmahon2016fully,
	aonishi2017statistical,
	aonishi2018critical,
	aonishi2018cdma,
	hamerly2019experimental}
Goto proposed a quantum adiabatic computation algorithm based on a nonlinear oscillator network that searches for the ground state of an Ising model. \cite{goto2016bifurcation,goto2019quantum}
This algorithm is implemented as a superconducting circuit \cite{nigg2017robust} or two-photon-driven Kerr parametric oscillators. \cite{puri2017quantum}
Goto also proposed a simulated bifurcation algorithm,
which is a classical approximation of a quantum adiabatic computation using a nonlinear oscillator network,
implemented in field-programmable gate arrays (FPGAs). \cite{goto2019combinatorial}
Other examples of such machines include electromechanical resonators, \cite{Mahboobe1600236}
nano-magnet arrays, \cite{sutton2017intrinsic}
electronic oscillators, \cite{parihar2017vertex}
and laser networks. \cite{tait2017neuromorphic}
There are machines based on simulated annealing (SA),
implemented in complementary metal-oxide-semiconductor (CMOS), \cite{
	yamaoka201520k,
	yoshimura2015uncertain,
	okuyama2016computing,
	zhang2017advancing
}
FPGAs, \cite{
	yoshimura2016fpga,
	yoshimura2017implementation,
	tsukamoto2017accelerator,
	sao2019application,
	aramon2019physics
}
and magnetic devices. \cite{mizushima2017large}

Many of these Ising computers have hardware restrictions on the implementation of their algorithm.
For example,
the superconducting quantum annealing processor \cite{
	johnson2011quantum,
	bunyk2014architectural
}
restricts the graph topology
to a chimera graph.
CMOS annealing \cite{
	yamaoka201520k,
	yoshimura2015uncertain
}
restricts the graph topology to a three-dimensional lattice built from two-layer two-dimensional lattices.
Direct mapping of most of the combinatorial optimization problems onto Ising models requires all-to-all couplings.
Thus, we have to translate the all-to-all coupling Ising models
into equivalent Ising models with other graph topologies
implementable on these machines.
Some translation techniques have been proposed. \cite{
	choi2008minor,
	choi2011minor,
	lechner2015quantum,
	albash2016simulated
}
As other examples,
the measurement-feedback type of CIM \cite{
	inagaki2016coherent,
	mcmahon2016fully,
	aonishi2017statistical,
	aonishi2018critical,
	aonishi2018cdma,
	hamerly2019experimental
}
and the Digital Annealer \cite{tsukamoto2017accelerator}
require that the coupling strength takes a discrete value.
Because calculating the local field requires large computing resources of digital circuits (\eg~FPGA),
the number of bits representing the coupling strength is the main factor determining
the number of implemented spins, processing speed, and development cost.
The number of bits representing the coupling strength should be made as small as possible
while maintaining performance as much as possible.

\begin{figure*}[t]
	\centering
	\includegraphics[width=0.9\textwidth]{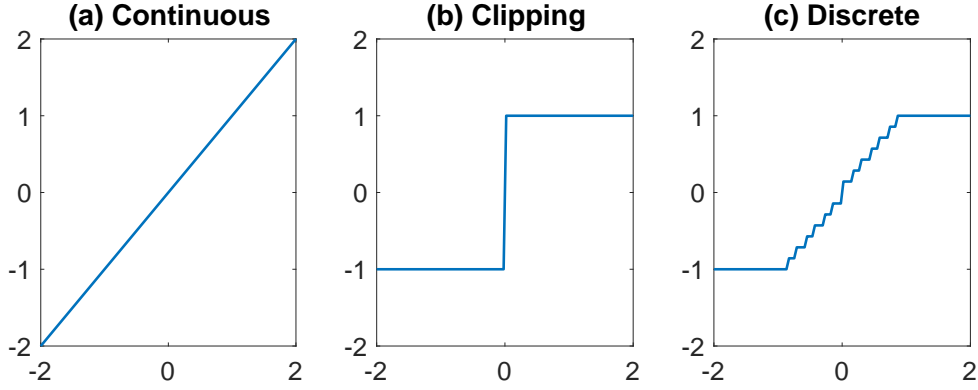}
	\caption{
		Examples of the discretization function $f$.
		(a): Linear function $f(x)=x$, which is used in the definition of the original Hopfield model.
		(b): Signum function $f(x)=\sgn(x)$, which is used in two-bit coupling strength, called clipping synapses.
		(c): Multi-bit discretization function defined as \eqref{eq:multibits_system}
		in the case of $n=4$.
	}
	\label{fig:functions}
\end{figure*}

Therefore, there is a growing demand for evaluating the effect of such hardware restrictions on the performance of Ising computers.
In this paper,
we focus on the Hopfield model with discrete coupling. \cite{
	hopfield1982neural, 
	sompolinsky1986neural, 
	sompolinsky1987theory, 
	mimura1996,
	okada1998random
}
The Hopfield model with a two-bit coupling strength,
named ``clipping synapses'', 
was analyzed using the replica method \cite{sompolinsky1986neural}
and self-consistent signal-to-noise analysis (SCSNA). \cite{okada1998random}
Moreover, the perceptron with an up-to-four-bit discrete coupling strength has been analyzed. \cite{gutfreund1990capacity, nokura1994capacity}
On the other hand,
Mimura \etal~have analyzed the Hopfield model with a multi-bit discrete coupling strength,
in which discrete intervals were non-uniformly optimized to maximize its memory capacity. \cite{mimura1996}
However, the multi-bit discretization manner proposed by Mimura \etal
is different from the practical manner of integer and fixed-point representations
used in the Ising-computing-specific systems developed recently.
Thus, the evaluation of the systems with the practical multi-bit discretization manner for coupling strength is demanded.

The Hopfield model shares many statistical mechanics pictures 
with other Ising models.
Therefore,
through an analysis of the Hopfield model with the practical multi-bit discretization manner for a coupling strength,
we expect to be able to estimate how many bits are needed
to represent coupling strengths
and at the same time
speculate the performance of other Ising models with such discrete couplings.
In this paper,
we used the replica method for the Hopfield model with general nonlinear coupling \cite{sompolinsky1986neural,sompolinsky1987theory}
to analyze the model with a multi-bit discrete coupling strength,
and we novelly derived the de Almeida-Thouless (AT) line of the model with general nonlinear coupling.
Moreover, as mentioned above,
there is a novelty
that we theoretically evaluate the performance of Ising-computing-specific systems 
with the practical discrete representation for the coupling strength.

\section{Model}

In the original Hopfield model,
the coupling strength was determined using the Hebb rule. \cite{hebb1949organization,hopfield1982neural}
In this study, 
we determined the coupling strength restricted to discrete values
by using the following modified Hebb learning rule, 
\cite{sompolinsky1986neural,okada1998random}
\begin{align}
	J_{ij} =& \cfrac{\sqrt{p}}{N} f(T_{ij}), 
		\label{eq:connectivity_modified_hebb} \\
		T_{ij} =& \left\{ \begin{matrix}
			\displaystyle \cfrac{1}{\sqrt{p}} \sum_{\mu=1}^p \xi_i^{\mu} \xi_j^{\mu}, & (i \neq j) \\
			0 & (i = j),
		\end{matrix} \right.  \notag
\end{align}
where
$\bg{\xi}^{\mu} = (\xi_1^{\mu},\dots,\xi^{\mu}_N)^T \in \{1, -1\}^N$ is the $\mu$-th memory pattern,
$p$ is the number of patterns,
$N$ is the system size,
and $f$ is a function to discretize the coupling strength.
The memory patterns are generated according to the probability distribution,
\begin{align}
	\Pr [\xi_i^{\mu} = \pm 1] = \cfrac{1}{2}.
\end{align}
The original Hopfield model corresponds to having a linear function $f(x) = x$.
The two-bit coupling strength, called clipping synapse, is determined using the signum function $f(x) = \sgn(x)$.
To determine the multi-bit coupling strength,
we define the discretization function $f$ as follows:
\begin{align}
	f(x) =&
		\begin{cases}
			\cfrac{\lfloor (2^{n-1}-1)x \rfloor}{2^{n-1} - 1} & (-1 < x < 0) \\
			\cfrac{\lceil (2^{n-1}-1)x \rceil}{2^{n-1} - 1} & (0 \leq x < 1) \\
			\sgn (x) & (|x| \geq 1)
		\end{cases}
		\label{eq:multibits_system}
		,
\end{align}
where $n$ represents the number of bits,
$\lfloor \cdots \rfloor$ represents the floor function,
and $\lceil \cdots \rceil$ represents the ceil function.
In addition, we introduce a loading rate $\alpha$, defined as $\alpha = p / N$,
and a local field $h_i$ at the $i$-th site, defined as
\begin{align}
	h_i = \sum_{j=1}^N J_{ij} S_j.
	\label{eq:local_field_definition}
\end{align}

\eqrefInit{eq:multibits_system}
discretizes the coupling strength in the range of $-1$ to $1$.
We attempted to verify how the phase diagram changes
as the range of the discretization function changes.
Thus, we modified \eqref{eq:connectivity_modified_hebb} as follows:
\begin{align}
	J_{ij} = \cfrac{\sqrt{p}}{N} g(T_{ij}), \quad
	g(x) = n_{\sigma} f(x / n_{\sigma}), 
	\label{eq:nonlinear_learning_with_sigma}
\end{align}
where $n_{\sigma}$ is a parameter which decides the range of the discretization function.
$n_{\sigma} f(x/n_{\sigma})$ in \eqref{eq:nonlinear_learning_with_sigma} 
discretizes the value of $x$ in the range of $-n_{\sigma}$ to $n_{\sigma}$.
For example, when $n_{\sigma}=2$, this function discretizes values in the range of $-2$ to $2$.

\figrefInit{fig:functions} shows the profiles of the discretization functions.
\figrefInit{fig:functions} (a) shows the linear function, which is used in the definition of the original Hopfield model.
Figures \ref{fig:functions} (b) and (c) indicate the functions to discretize the coupling strength
into two-bit and multi-bit values, respectively.

\section{Theory}

\subsection{Hebbian-Glassy Coupling Effectively Equivalent to Discretized Coupling}

As a first step,
by performing a naive signal-to-noise (S/N) analysis,
we derive a Hebbian-glassy coupling effectively equivalent to \eqref{eq:nonlinear_learning_with_sigma}.
When $S_i = \xi_i^{\nu}$,
the local field \eqref{eq:local_field_definition} is
\begin{align}
	h_i =& \cfrac{\sqrt{p}}{N} \sum_{j=1}^N g (T_{ij}) S_j \notag \\
	=& \cfrac{1}{N} \sum_{j \neq i}^N \xi_i^{\nu} \xi_j^{\nu} \xi_j^{\nu} g^{\prime} (T_{ij}^{(\nu)})
		+ \cfrac{\sqrt{p}}{N} \sum_{j \neq i}^N g(T_{ij}^{(\nu)}) \xi_j^{\nu}, 
		\label{eq:local_field_SN} \\
	T_{ij}^{(\nu)} =& \cfrac{1}{\sqrt{p}} \sum_{\mu\neq\nu}^p \xi_i^{\mu} \xi_j^{\mu}.
		\label{eq:covariance_learning_without_condensed_pattern}
\end{align}
The first part of \eqref{eq:local_field_SN} is the signal term,
and the second part is the noise term.
In the limit $N \rightarrow \infty$, the signal term can be rewritten as 
\begin{align}
	&\cfrac{1}{N} \sum_{j \neq i}^N \xi_i^{\nu} \xi_j^{\nu} \xi_j^{\nu} g^{\prime}(T_{ij}^{(\nu)}) 
	= J\xi_i^{\nu}, 
		\label{eq:sn_signal} \\
	& J \equiv \int Dx g^{\prime}(x)
		= \int Dxx g(x),
		\label{eq:definition_J}
\end{align}
where $Dx = dx e^{-x^2/2} / \sqrt{2\pi} $.
Because $\xi^{\nu}_i$ and $T_{ij}^{(\nu)}$ are independent
and $T_{ij}^{(\nu)}$ obeys a Gaussian distribution by the central limit theorem,
\seqeqref{eq:sn_signal}{eq:definition_J} can be obtained.
On the other hand,
the mean and the variance of the noise term become
\begin{align}
	&\llangle \cfrac{\sqrt{p}}{N} \sum_{j \neq i}^N g(T_{ij}^{(\nu)}) \xi_j^{\nu} \rrangle = 0, \\
	&\llangle \left(
		\cfrac{\sqrt{p}}{N} \sum_{j \neq i}^N g(T_{ij}^{(\nu)}) \xi_j^{\nu}
	\right)^2 \rrangle = \alpha \tilde{J}, \\
	&\tilde{J} = \int Dx g(x)^2,
	\label{eq:definition_tilJ}
\end{align}
where $\llangle\cdots\rrangle$ implies averaging over all of the random memory patterns $\{\xi_i^{\mu}\}$.
The result of the S/N analysis for the pattern $\xi^k_i$ are satisfied for any $k=1, \dots, N$.

By adding and subtracting the same term $J\sqrt{p}T_{ij}/N$ to/from \eqref{eq:nonlinear_learning_with_sigma},
the coupling strength $J_{ij}$ defined in \eqref{eq:nonlinear_learning_with_sigma}
can be rewritten as follows,
\begin{align}
	J_{ij} = \cfrac{J\sqrt{p}}{N} T_{ij} 
		+ \cfrac{\sqrt{p}}{N} (g(T_{ij}) - JT_{ij}).
		\label{eq:snanalysis}
\end{align}
The first and second parts of \eqref{eq:snanalysis}
correspond to the signal term and noise term in \eqref{eq:local_field_SN}, respectively.
Assuming that a signal condensed pattern exists,
the first and second parts in \eqref{eq:snanalysis} can be considered to be statistically independent.
Thus, according to the central limit theorem,
the second part in \eqref{eq:snanalysis} can be replaced by a Gaussian random variable
with zero mean and variance $\alpha(\tilde{J} - J^2)/N$
in the limit of $N \rightarrow \infty$.
Thus, we get
\begin{align}
	J_{ij} = \cfrac{J\sqrt{p}}{N} T_{ij} + \eta_{ij}
		= \cfrac{J}{N} \sum_{\mu=1}^p \xi_i^{\mu} \xi_j^{\mu} + \eta_{ij},
		\label{eq:connectivity_with_static_noise}
\end{align}
	where the glassy coupling part $\eta_{ij} (i \neq j)$ has been proved to be an independently and identically distributed Gaussian random variable
	with zero mean and variance $J^2\Delta^2/N$
	independent of the Hebbian rule part $J/N \sum_{\mu=1}^p \xi_i^{\mu} \xi_j^{\mu}$. \cite{sompolinsky1986neural}
Note that $\eta_{ij} = \eta_{ji}$ (symmetry) and $J_{ii} = \eta_{ii} = 0$.
$\Delta$ is defined as
\begin{align}
	\Delta^2 = \alpha \left( \cfrac{\tilde{J}}{J^2} - 1 \right).
\end{align}

\subsection{Replica Method}

In this subsection, we analyze using the replica method, 
following the recipe of the previous study. \cite{%
	amit1987statistical,%
	sompolinsky1986neural%
}%
We introduce the temperature $T = \beta^{-1}$ %
and define the partition function as follows:
\begin{align}
	Z = \Tr \exp (-\beta \ham).
	\label{eq:partition_function}
\end{align}
Applying the replica trick,
we derive the average free energy per spin $f$.
The details are given in \appnref{sec:appendix_free_energy}.
Assuming replica symmetric theory,
we obtain the following equation.

\begin{align}
	f
	=& - \lim_{n\rightarrow0}\lim_{N\rightarrow\infty} \cfrac{\llangle [Z^n] \rrangle -1}{\beta n N} \notag \\
	=& \cfrac{J\alpha}{2} 
		- \cfrac{\alpha}{2\beta} \left\{
			\log (1-J\beta+J\beta q)
			- \cfrac{J\beta q}{1-J\beta+J\beta q}
		\right\}
		\notag \\
	 &+ \cfrac{J}{2} \left\{
			J\alpha\beta r(1-q) + \left(m^1\right)^2
		\right\}
	- \cfrac{J^2\beta\Delta^2(1-q)^2}{4} \notag \\
	& - \cfrac{1}{\beta} \llangle \int Dz
		\log 2 \cosh J\beta (\sqrt{\alpha r + \Delta^2 q} z + m^1\xi^1 )
	\rrangle_{\xi^1}.
	\label{eq:free_energy_rs}
\end{align}

\begin{figure*}[t]
	\def\xpos{105}
	\def\ypos{80}
	\def\atwid{4cm}
	\subfigure[]{
		\begin{overpic}[width=0.48\textwidth,clip]{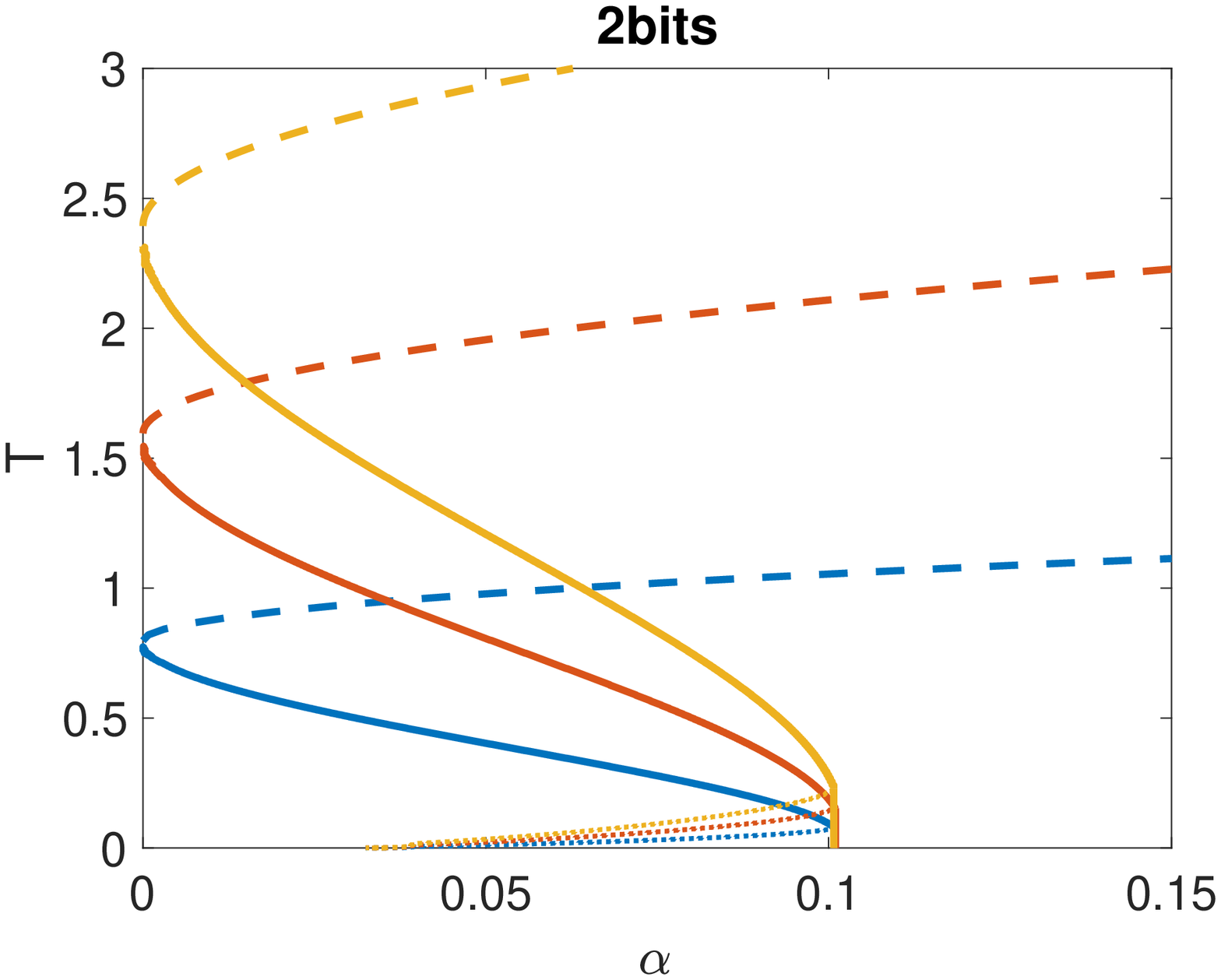}
			\put (\xpos,\ypos) {\includegraphics[width=\atwid]{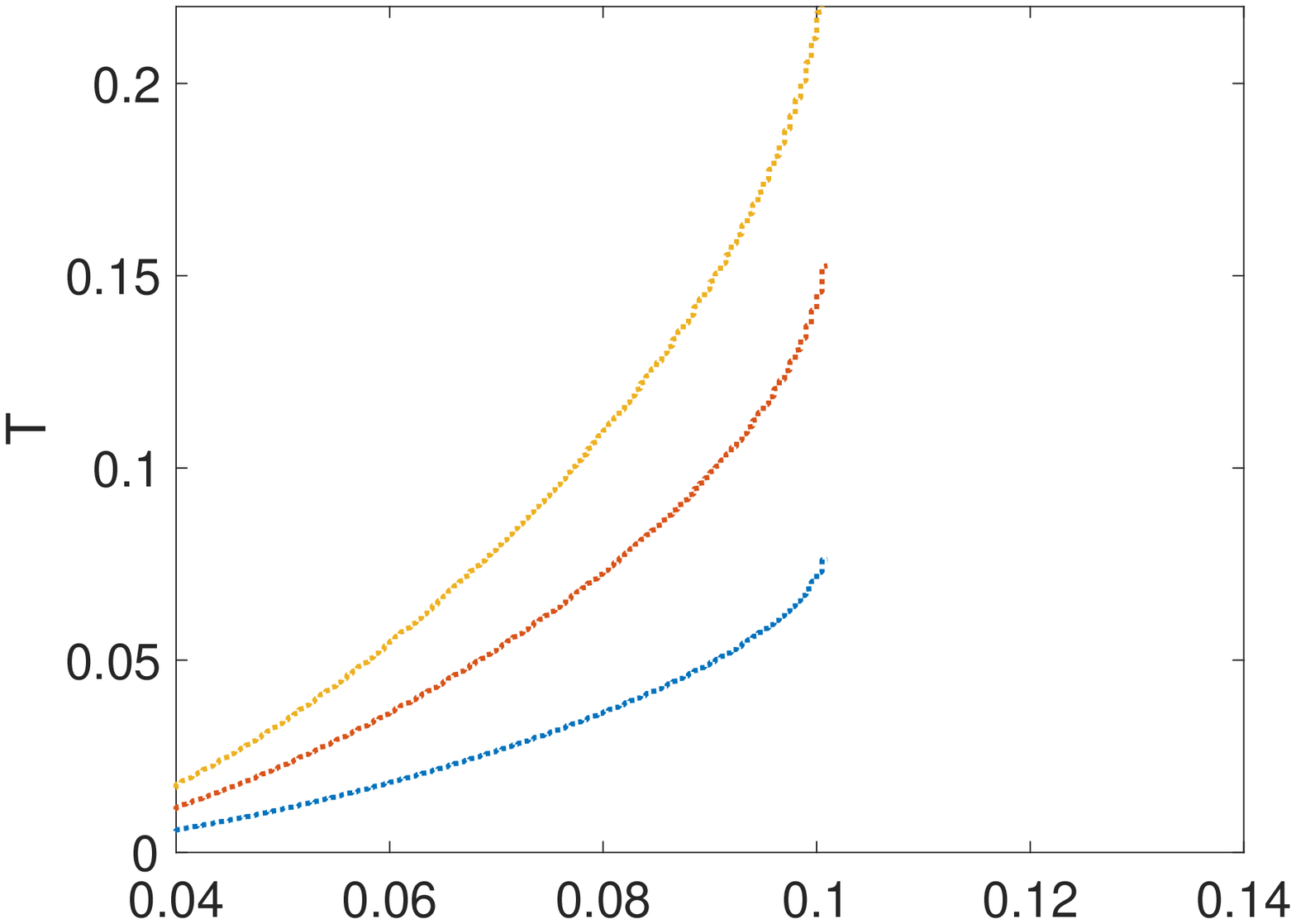}}
		\end{overpic}
		\label{fig:subfig:replica_2bit}
	}
	\subfigure[]{
		\begin{overpic}[width=0.48\textwidth,clip]{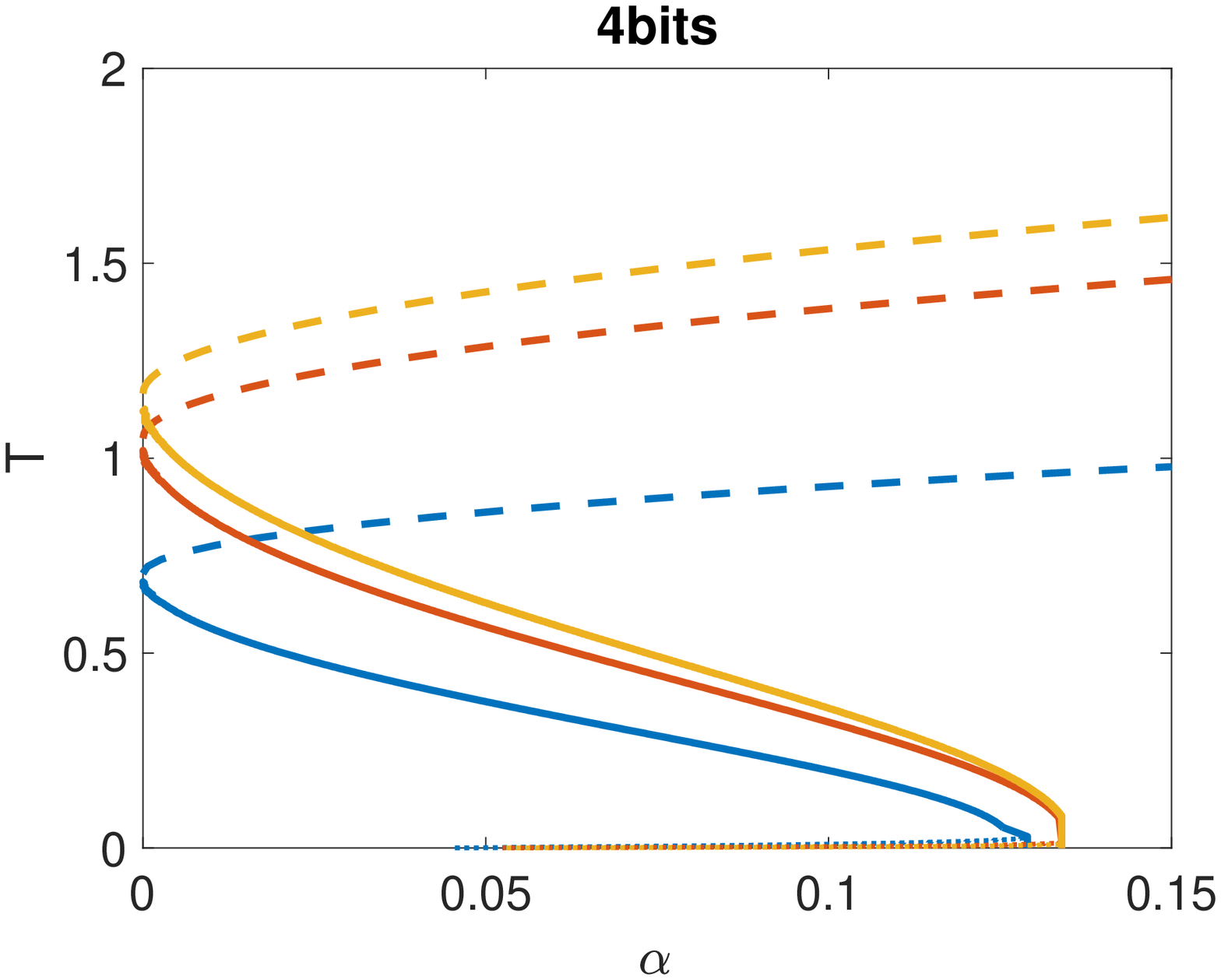}
			\put (\xpos,\ypos) {\includegraphics[width=\atwid]{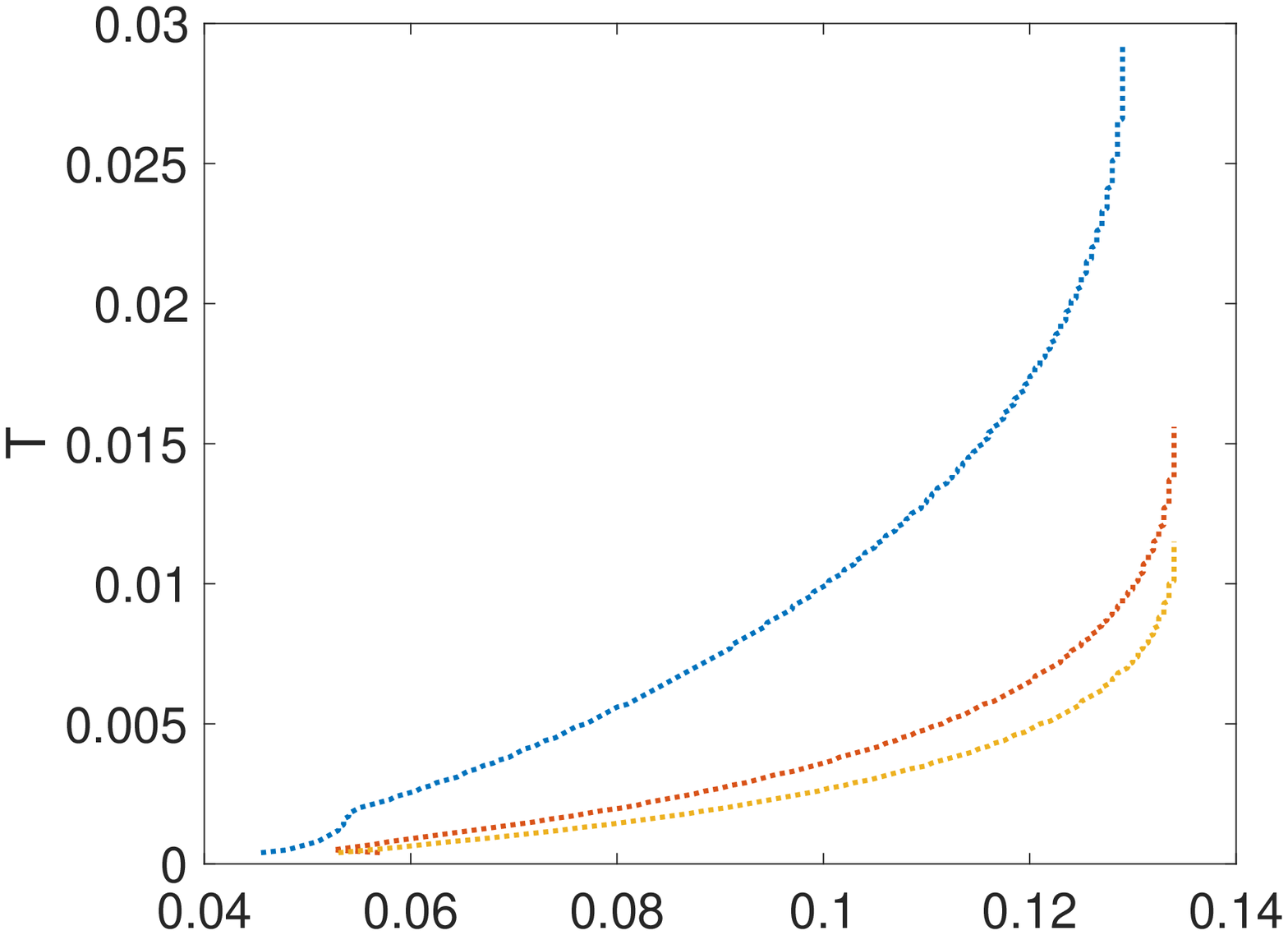}}
		\end{overpic}
		\label{fig:subfig:replica_4bit}
	}
	\subfigure[]{
		\begin{overpic}[width=0.48\textwidth]{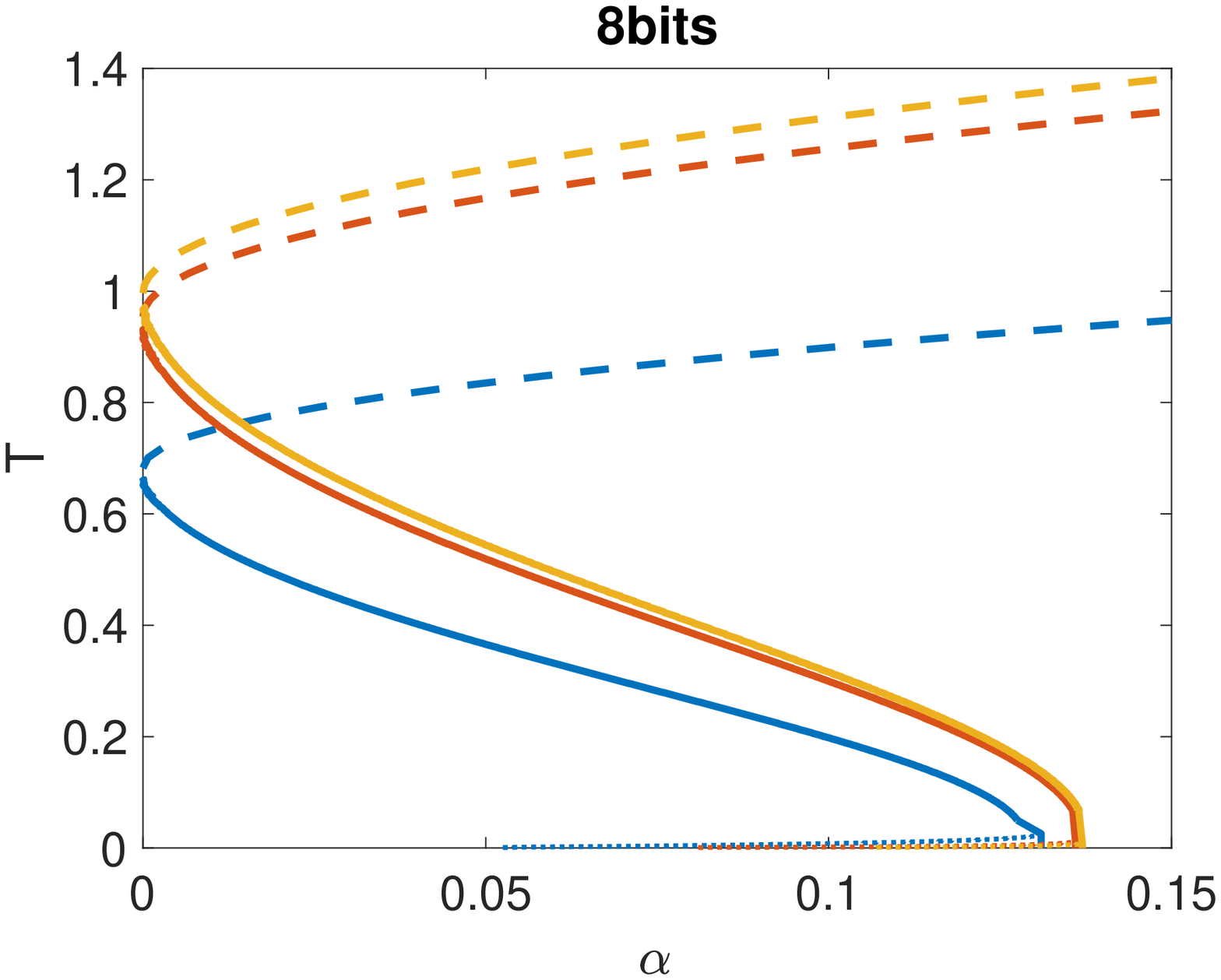}
			\put (\xpos,\ypos) {\includegraphics[width=\atwid]{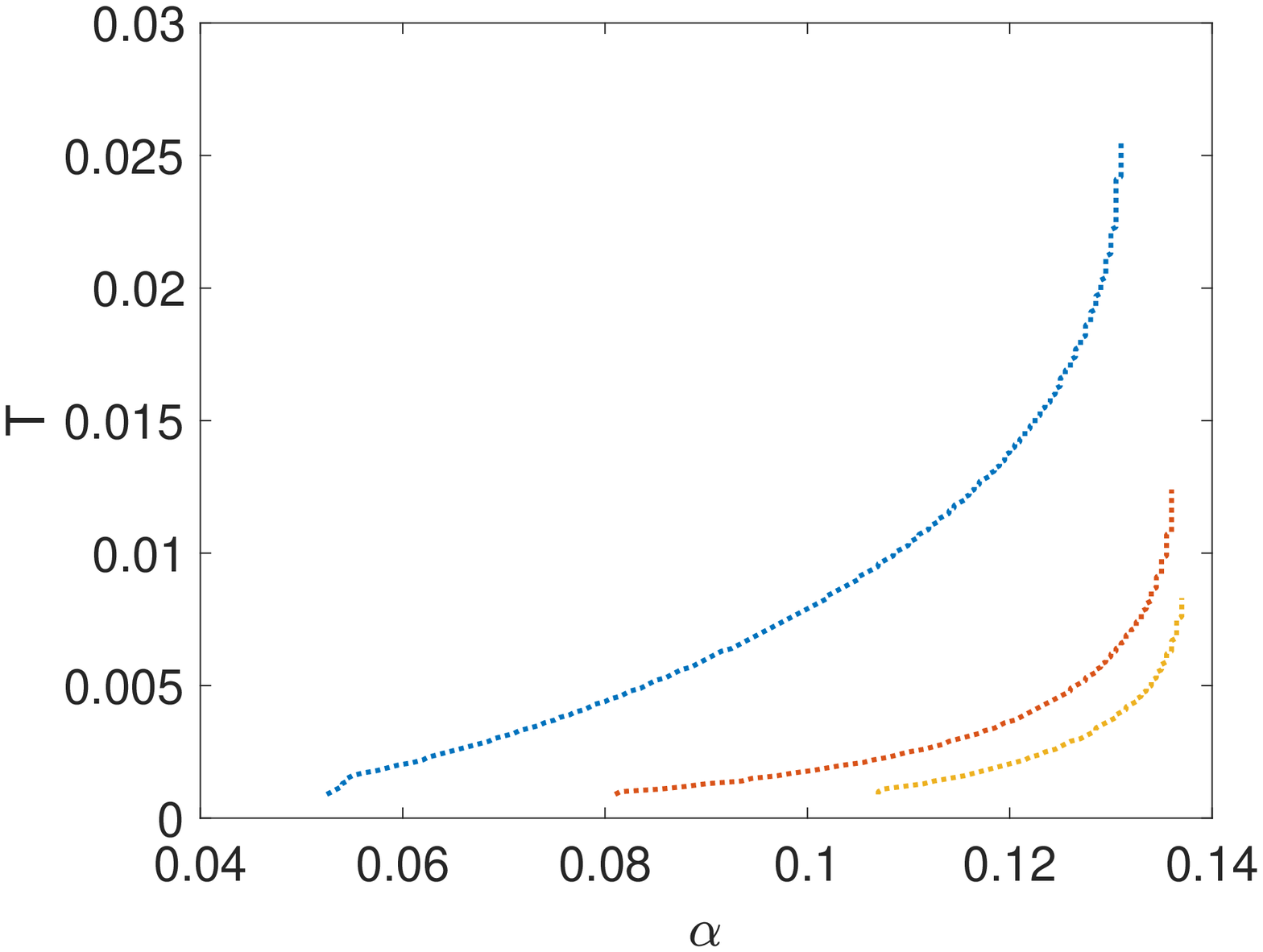}}
		\end{overpic}
		\label{fig:subfig:replica_8bit}
	}
	\begin{minipage}[t]{0.45\textwidth}
		\centering
		\includegraphics[width=0.8\textwidth]{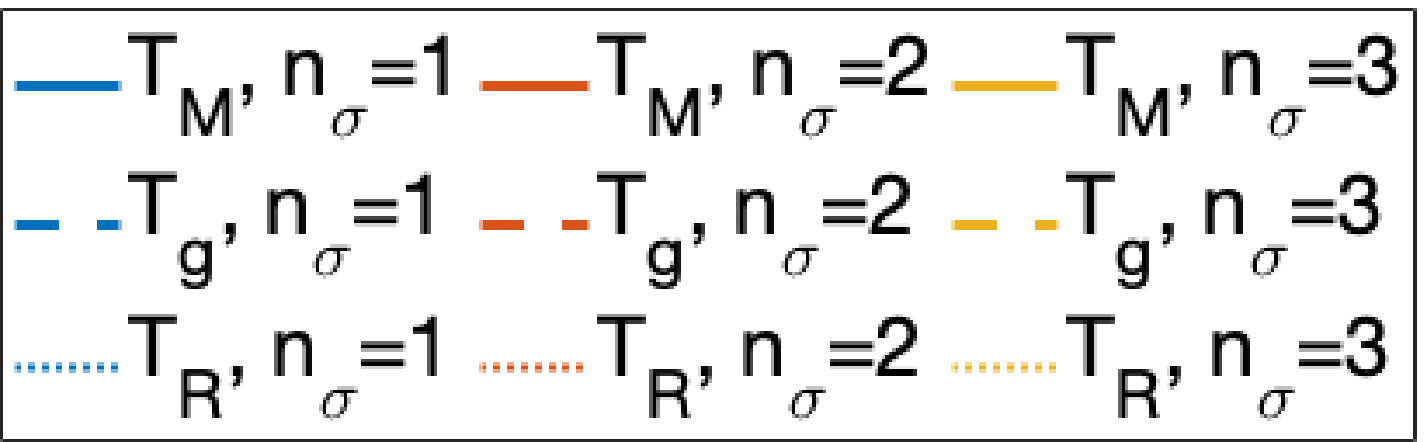}
	\end{minipage}
	\caption{
		Plots of critical temperatures of the SG and the FM states as a function of $\alpha$.
		(a): Case of two-bit coupling strength.
		(b): Case of four-bit coupling strength.
		(c): Case of eight-bit coupling strength.
		In each panel, the range of the discretization function varies as $n_{\sigma}=1,2,3$.
		The dashed line shows the transition temperature $T_g$ to the SG state.
		The solid line shows the temperature $T_M$ at which the FM states first appear.
		Replica symmetry is broken below the dotted line $T_R$.
	}
	\label{fig:replica}
	\let\xpos=\free
	\let\ypos=\free
	\let\atwid=\free
\end{figure*}

Here, $[\cdots]$ implies averaging over the glassy coupling part
and $\llangle\cdots\rrangle_{\xi^1}$ denotes averaging over the random pattern $\xi^1$.
$m^1$ is an order parameter called the macroscopic overlap,
defined as the correlation between a state of spins and a condensed pattern $\{\xi_i^1\}$,
\begin{align}
	m^1 = \llangle \left[ \cfrac{1}{N} \sum_{i=1}^N \xi_i^1 \langle S_i \rangle_T \right] \rrangle,
\end{align}
where $\langle \cdots \rangle_T$ represents the thermal average.
$q$ is the Edwards-Anderson order parameter,
\begin{align}
	q = \llangle \left[ \cfrac{1}{N} \sum_{i=1}^N \langle S_i \rangle_T^2 \right] \rrangle.
\end{align}
$r$ is the mean-square of the overlaps with uncondensed patterns,
\begin{align}
	r = \cfrac{1}{\alpha} \llangle \left[ \sum_{\mu=2}^p \left( \cfrac{1}{N} \sum_{i=1}^N \xi_i^{\mu} \langle S_i \rangle_T \right)^2 \right] \rrangle.
\end{align}
Extremizing $f$ with respect to $q$, $r$, and $m^1$,
we obtain the following saddle-point equations.
\begin{subequations}
\begin{align}
		m^1 =& \int Dz \tanh J\beta (\sqrt{\alpha r + \Delta^2 q}z + m^1), \\
		q =& \int Dz \tanh^2 J\beta (\sqrt{\alpha r + \Delta^2 q}z + m^1), \\
		r =& \cfrac{q}{(1-J\beta+J\beta q)^2}.
\end{align}
	\label{eq:suddle_point_equations}
\end{subequations}

\begin{figure*}[t]
	\subfigure[]{
		\includegraphics[width=0.32\textwidth]{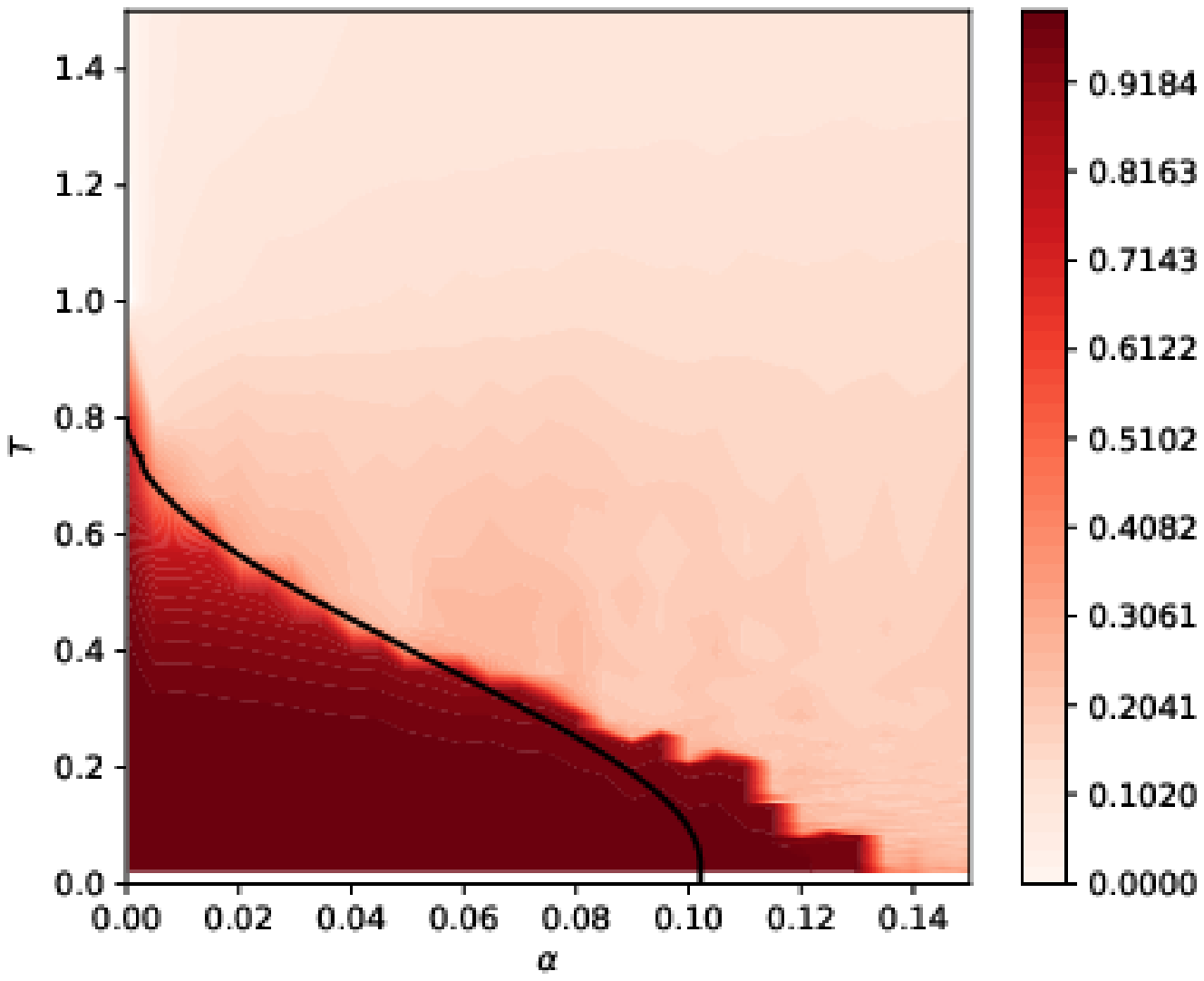}
		\label{fig:subfig:mcmc_replica_2bit}
	}
	\subfigure[]{
		\includegraphics[width=0.32\textwidth]{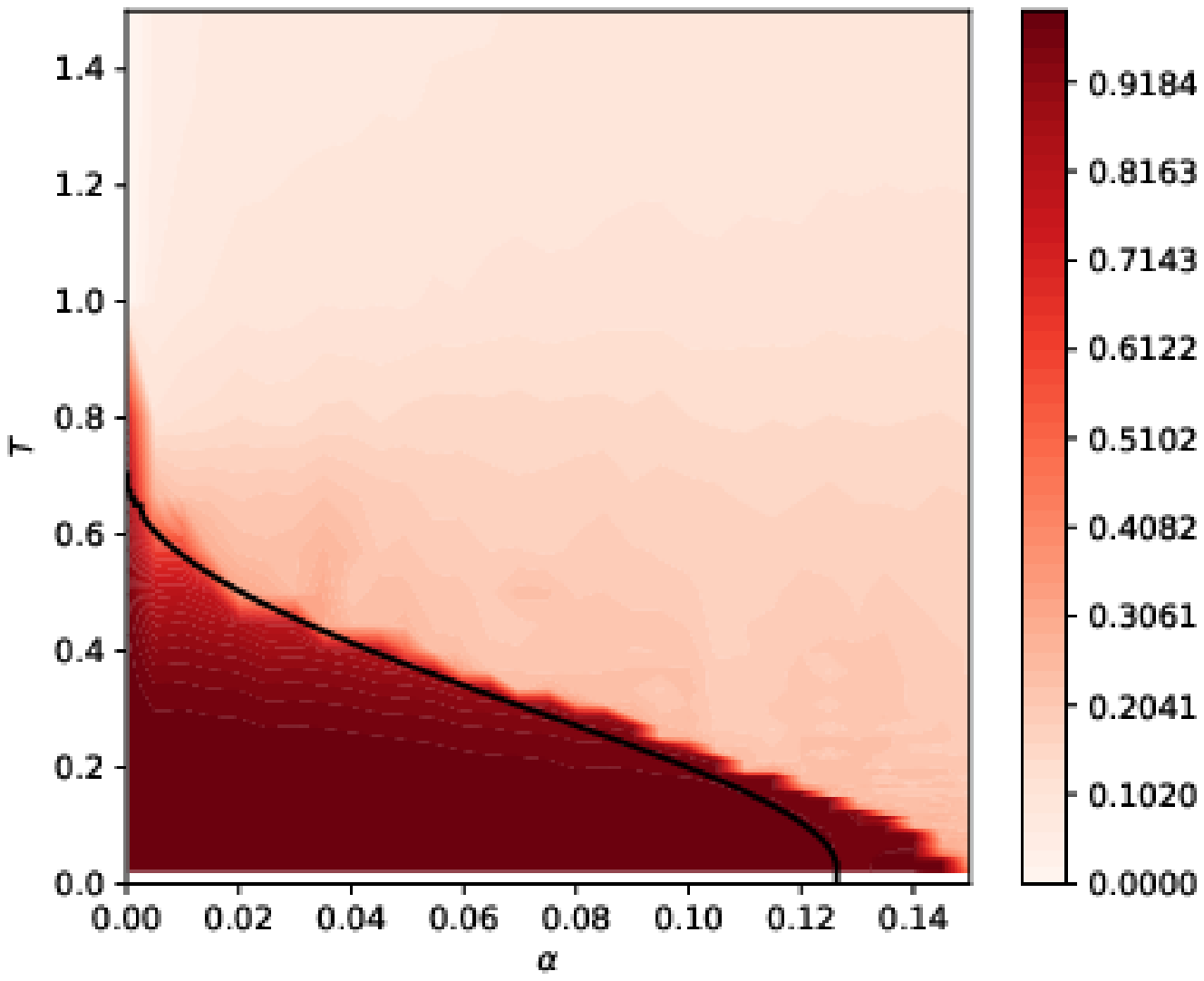}
		\label{fig:subfig:mcmc_replica_4bit}
	}
	\subfigure[]{
		\includegraphics[width=0.32\textwidth]{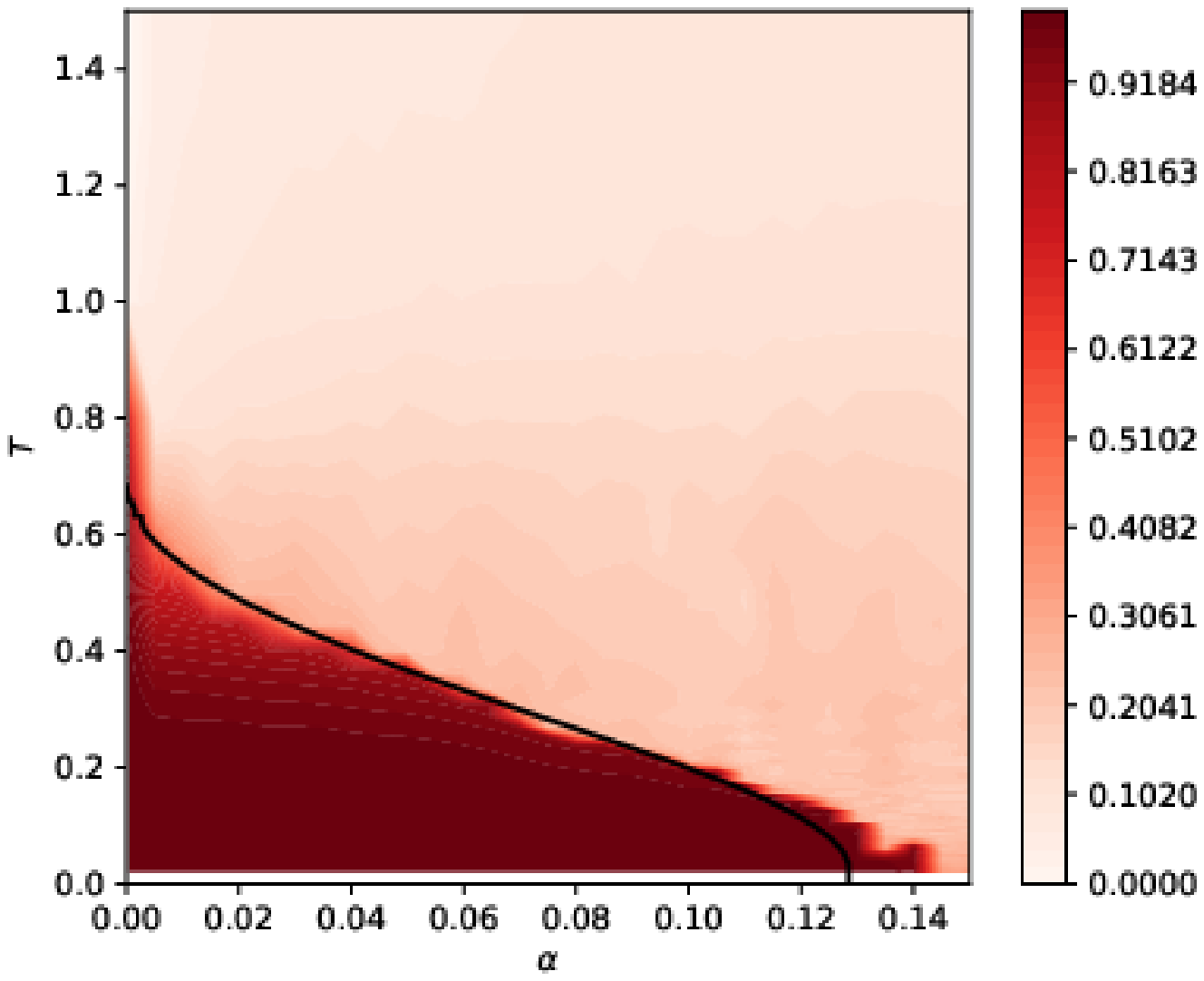}
		\label{fig:subfig:mcmc_replica_8bit}
	}
	\subfigure[]{
		\includegraphics[width=0.32\textwidth]{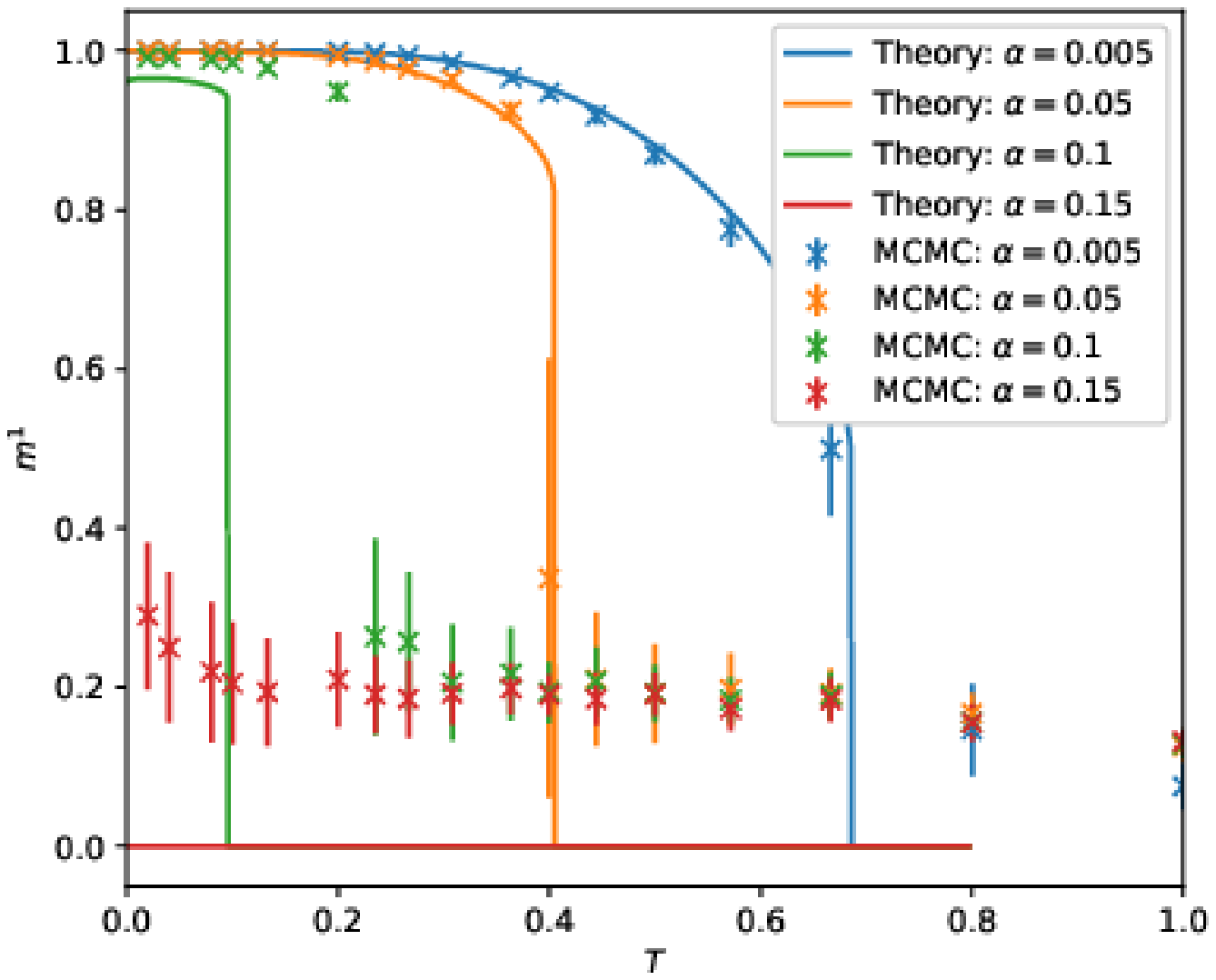}
		\label{fig:subfig:mcmc_replica_errorbar_2bit}
	}
	\subfigure[]{
		\includegraphics[width=0.32\textwidth]{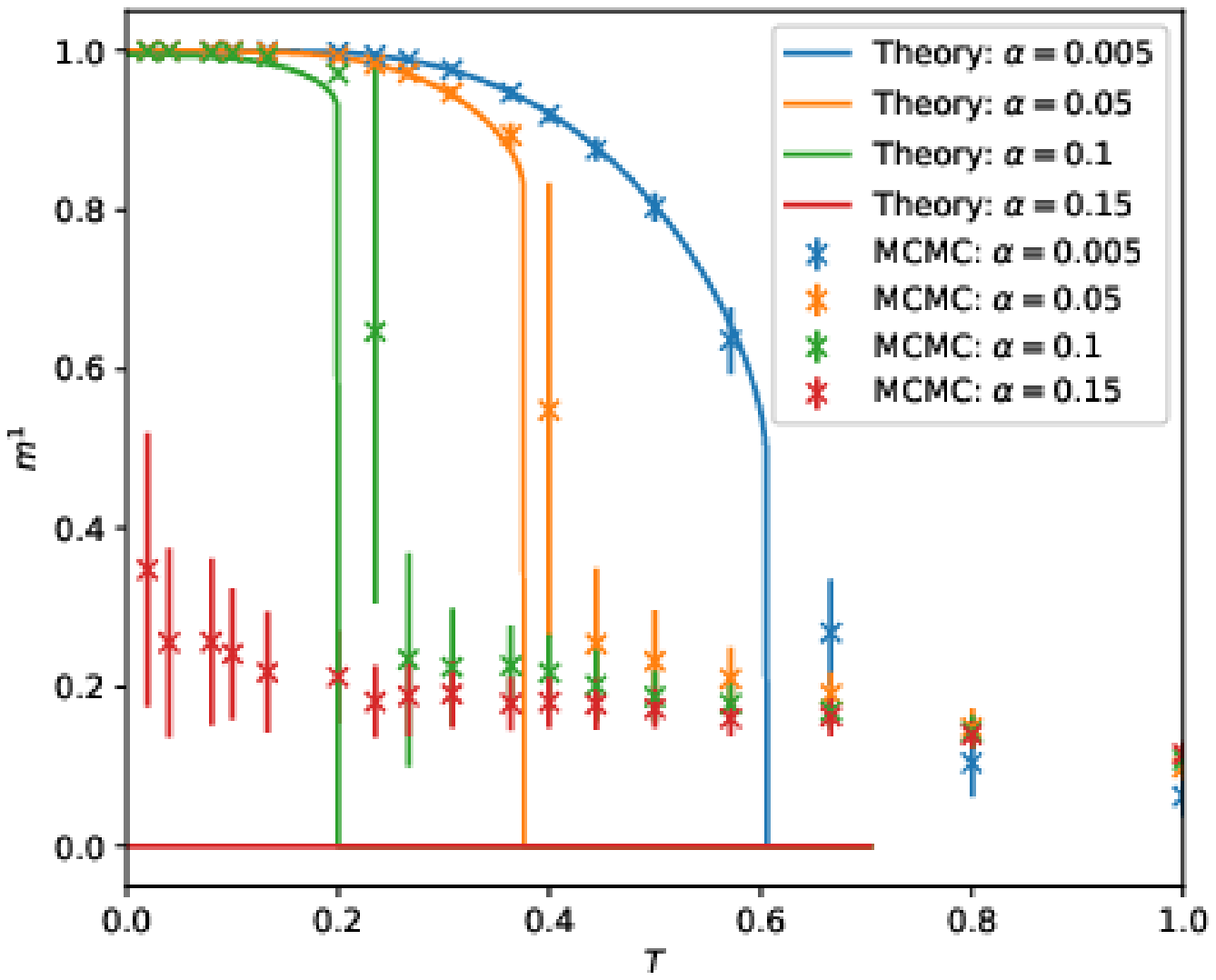}
		\label{fig:subfig:mcmc_replica_errorbar_4bit}
	}
	\subfigure[]{
		\includegraphics[width=0.32\textwidth]{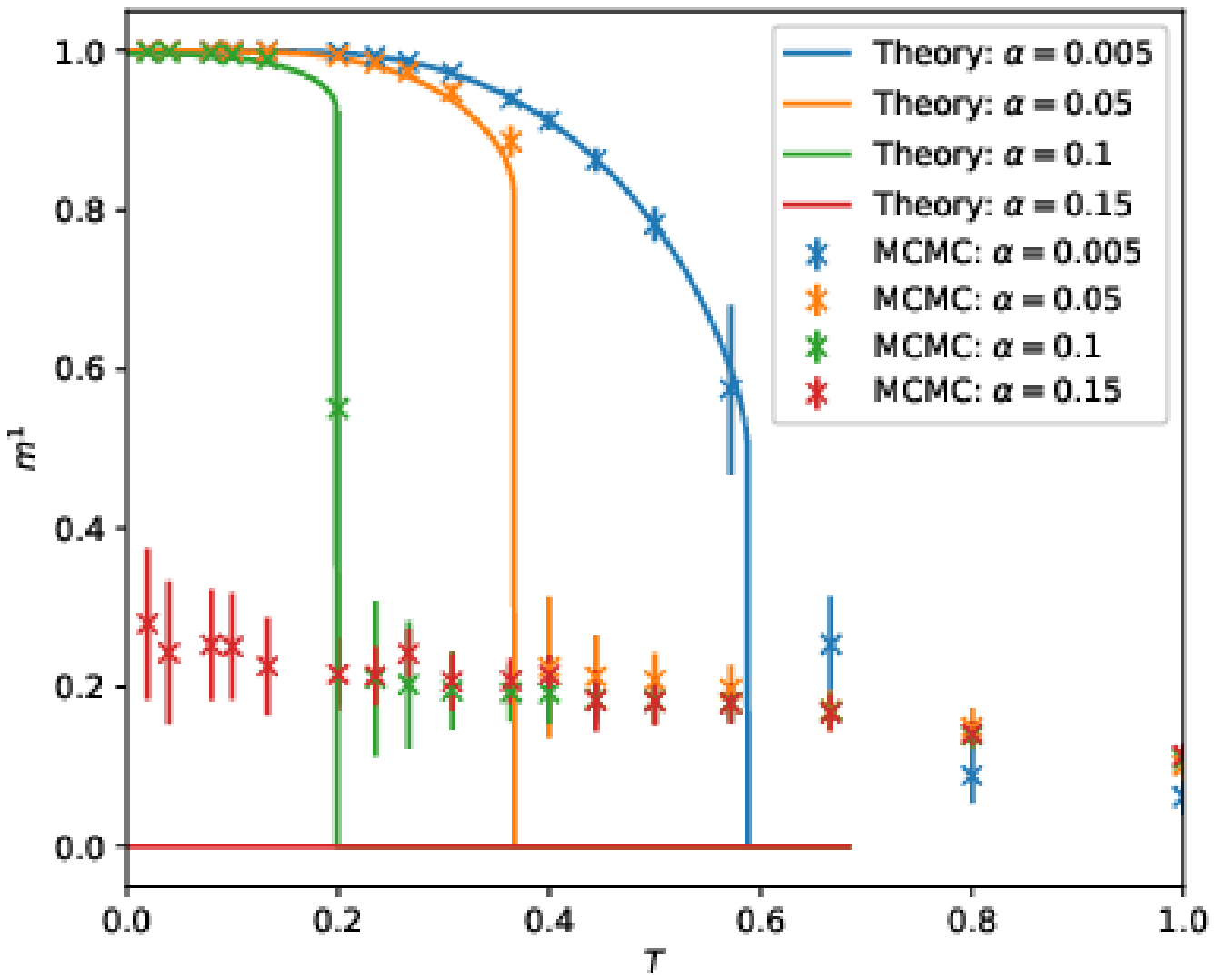}
		\label{fig:subfig:mcmc_replica_errorbar_8bit}
	}

	\caption{
		Plots of theoretical results obtained by saddle-point equations \eqref{eq:suddle_point_equations}
		and numerical results obtained by MCMC simulations.
		(a), (d): Case of two-bit coupling strength.
		(b), (e): Case of four-bit coupling strength.
		(c), (f): Case of eight-bit coupling strength.
		In (a)--(c), the solid line shows the transition temperature $T_M$ obtained by solving \eqref{eq:suddle_point_equations} numerically
		in the case of $n_{\sigma}=1$,
		and the color plot shows the value of $m^1$ obtained by MCMC simulations.
		In (d)--(f), the solid line shows the value of $m^1$ obtained by solving \eqref{eq:suddle_point_equations} numerically
		in the case of $n_{\sigma}=1$,
		and the symbols and error bars show means and standard deviations of the value of $m^1$ obtained by MCMC simulations.
		In (d)--(f), the loading rate $\alpha$ varies as $0.005, 0.05, 0.1$, and $0.15$.
	}
	\label{fig:mcmc_replica}
\end{figure*}

\eqrefInit{eq:suddle_point_equations} has a trivial solution $m^1 = q = r = 0$,
which is called the paramagnetic (PARA) phase.
Besides this phase,
there are two other phases.
One phase with $m^1\neq0,q\neq0$ is termed the ferromagnetic (FM) phase or retrieval phase.
The other phase with $m^1=0,q\neq0$ is termed the spin-glass (SG) phase.
Note that \eqref{eq:suddle_point_equations} is the same as the saddle point equations for the original Hopfield model in the case of  $J=1$ and $\tilde{J}=1$, resulting in $\Delta=0$. \cite{amit1987statistical}
\figrefInit{fig:replica} shows the phase diagram,
which plots the critical temperatures of the SG phase and the FM phase as a function of $\alpha$.
In each figure,
the dashed line shows the transition temperature $T_g$ to the SG phase,
the solid line shows the temperature $T_M$ at which the FM phase first appears,
and the dotted line shows the AT line.
We verified the cases of two-bit, four-bit and eight-bit coupling strengths to plot these transition temperatures for $n_{\sigma}=1, 2,$ and $3$.
The tricritical point of the FM phase, SG phase and PARA phase is $T=J$ and $\alpha=0$.

\subsubsection{SG Phase}

The transition from the PARA phase to the SG phase is of second order.
To find the transition temperature $T_g$,
we expand $q$ and $r$ in \eqref{eq:suddle_point_equations} under the assumption of a fixed $m^1 = 0$
and obtain a leading order equation,
\begin{align}
	q
	\approx J^2\beta^2 (\alpha r + \Delta^2 q)
	\approx J^2\beta^2 q \left\{
		\cfrac{\alpha + \Delta^2 (1-J\beta+J\beta q)^1}{(1-J\beta+J\beta q)^2} \right\},
	\label{eq:expand_q_around_tg}
\end{align}
which yields the following equation,
which determines the transition temperature $T_g$.
\begin{align}
	\alpha
	= (1-JT_g^{-1})^2 (J^{-2}T_g^2 - \Delta).
	\label{eq:tg_equation}
\end{align}

As shown in \figref{fig:subfig:replica_2bit}, 
$T_g$ increases as $J$ increases.
In the case of a two-bit coupling strength,
$J$ is proportional to $n_{\sigma}$, 
and $\Delta$ is constant with respect to $n_{\sigma}$.
Thus, $T_g$ obeying \eqref{eq:tg_equation} increases in proportion to $n_{\sigma}$.
On the other hand,
in the case of a four-bit or eight-bit coupling strength,
$T_g$ is no longer proportional to $n_{\sigma}$ (\seqfigref{fig:subfig:replica_4bit}{fig:subfig:replica_8bit}),
because both $J$ and $\Delta$ depend on $n_{\sigma}$.

For $T < T_g$, \eqref{eq:expand_q_around_tg} can be rewritten as
\begin{align}
	q
	\approx J^2\beta^2 (\alpha r + \Delta^2 q)
	\approx \cfrac{T(T_g-J)}{JT_g}
		\sqrt{\cfrac{T_g^2-J^2\Delta^2}{T^2-J^2\Delta^2}}
		- \cfrac{T}{J} + 1.
\end{align}
Thus, $q \neq 0$.

\subsubsection{FM Phase}

The FM phase is defined by $m^1 \neq 0$.
Above $T=J$, there are no FM solutions for any value of $\alpha$.
For $T<T_g$ and $\alpha<\alpha_c$, one finds the line $T_M(\alpha)$,
below which the FM phase appears.
Here, $\alpha_c$ is the critical memory capacity at $T \rightarrow 0$ (the details are described below).
In the FM phase, the macroscopic overlap $m^1$ becomes $\mathcal{O}(1)$,
which means retrieval of the condensed pattern $\{\xi_i^1\}$.
In the case of a two-bit coupling strength,
$T_M$ is proportional to $n_{\sigma}$, as is $T_g$.
In the case of four-bit and eight-bit coupling strengths,
$T_M$ increases with $n_{\sigma}$, but saturates for $n_{\sigma}$ larger than two for each case of $\alpha$.
Especially in the case of a four-bit coupling strength,
$T_M$ is maximized at about $n_{\sigma}=2$ and $\alpha>0.12$.
In each case,
as $\alpha$ approaches $\alpha_c$, the $T_M$ line asymptotically approaches the $T=0$ axis.

To confirm the accuracy of the saddle-point equations obtained by the replica method,
we performed the Markov Chain Monte Carlo (MCMC) simulation with Gibbs sampling for the system size $N=2,000$ 
in the case of $n_{\sigma}=1$.
\figrefInit{fig:mcmc_replica} shows the theoretical results obtained by saddle-point equations \eqref{eq:suddle_point_equations}
and the numerical results obtained by MCMC simulations.
In \figref{fig:subfig:mcmc_replica_2bit} -- \figref{fig:subfig:mcmc_replica_8bit},
the solid line in each subfigure shows the transition temperature $T_M$ obtained by solving \eqref{eq:suddle_point_equations} numerically,
and the color plot shows the value of $m^1$ obtained by MCMC simulations.
In \figref{fig:subfig:mcmc_replica_errorbar_2bit} -- \figref{fig:subfig:mcmc_replica_errorbar_8bit},
the solid lines in each subfigure show the values of $m^1$ as a function of $T$ with various $\alpha$,
which were obtained by solving \eqref{eq:suddle_point_equations} in the case of $n_\sigma=1$,
and the symbols and error bars show means and standard deviations of the values of $m^1$ obtained by MCMC simulations.
The phase transition points obtained from \eqref{eq:suddle_point_equations} coincided with those of the MCMC simulations
in the case of the two-bit, four-bit, and eight-bit coupling strengths
in many regions.
	However, as the number of bits was decreased and the loading rate $\alpha$ was increased,
	the transition temperature $T_M$ estimated with the saddle-point equation \eqref{eq:suddle_point_equations}
	was not matched to that with the MCMC simulations very well (see \seqfigref{fig:subfig:mcmc_replica_errorbar_2bit}{fig:subfig:mcmc_replica_4bit}).
	As suggested by Eqs. \ref{eq:definition_J}, \ref{eq:definition_tilJ}, and \eqref{eq:connectivity_with_static_noise},
	the relative strength of the effective glassy coupling part increases 
	as the number of bits decreases and $\alpha$ increases.
	Thus, we suspect that there might be many metastable states 
	due to the effective glassy coupling part,
	and thus,
	the relaxation time in the MCMC simulations might be longer.
	This discrepancy became more pronounced at lower temperatures (see \seqfigref{fig:subfig:mcmc_replica_2bit}{fig:subfig:mcmc_replica_4bit}),
	which supports the above suspicion.

\begin{figure*}[!t]
	\subfigure[]{
	\centering
	\includegraphics[width=0.49\textwidth]{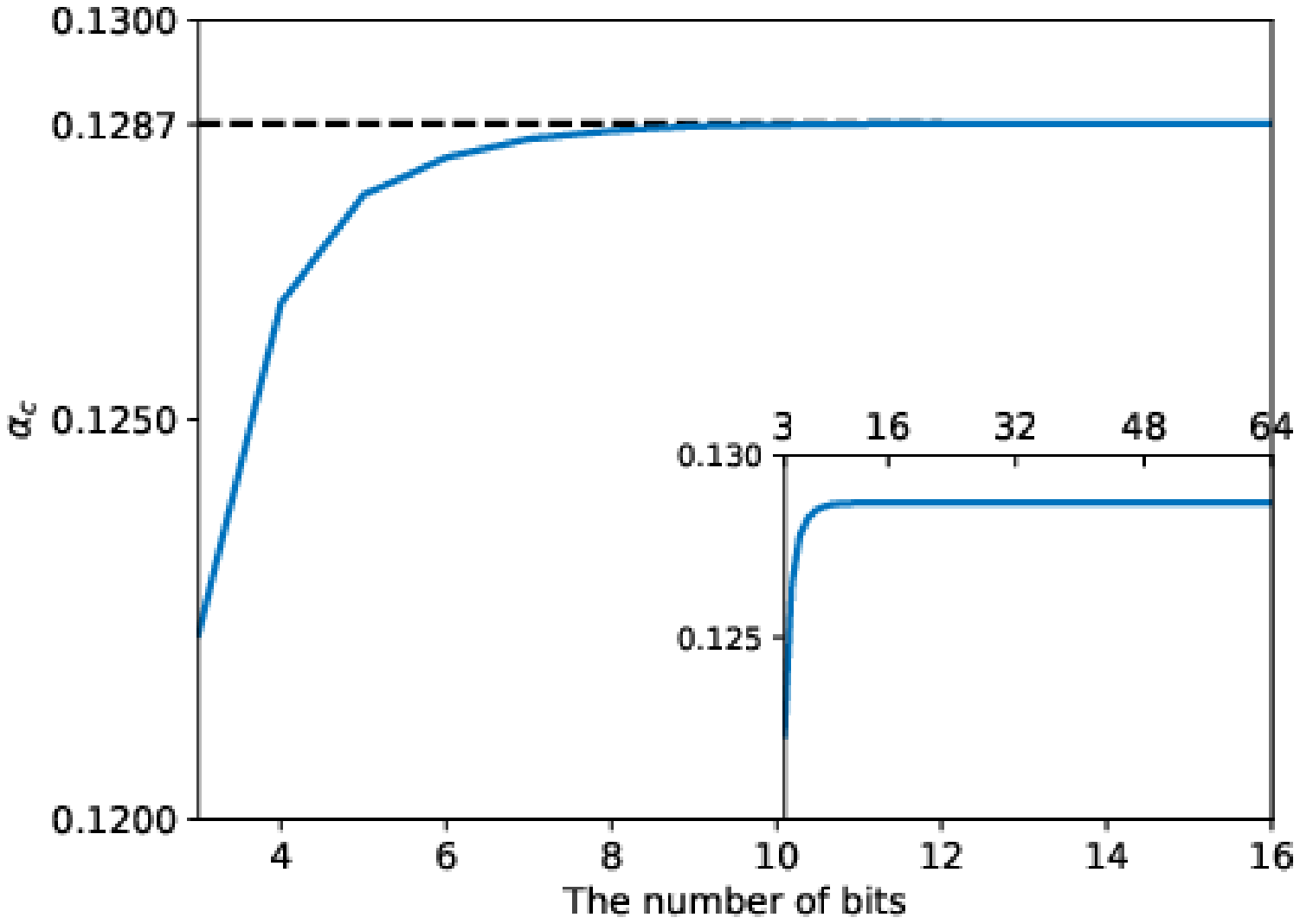}
	\label{fig:cmc_bits}
	}
	\subfigure[]{
	\centering
	\includegraphics[width=0.49\textwidth]{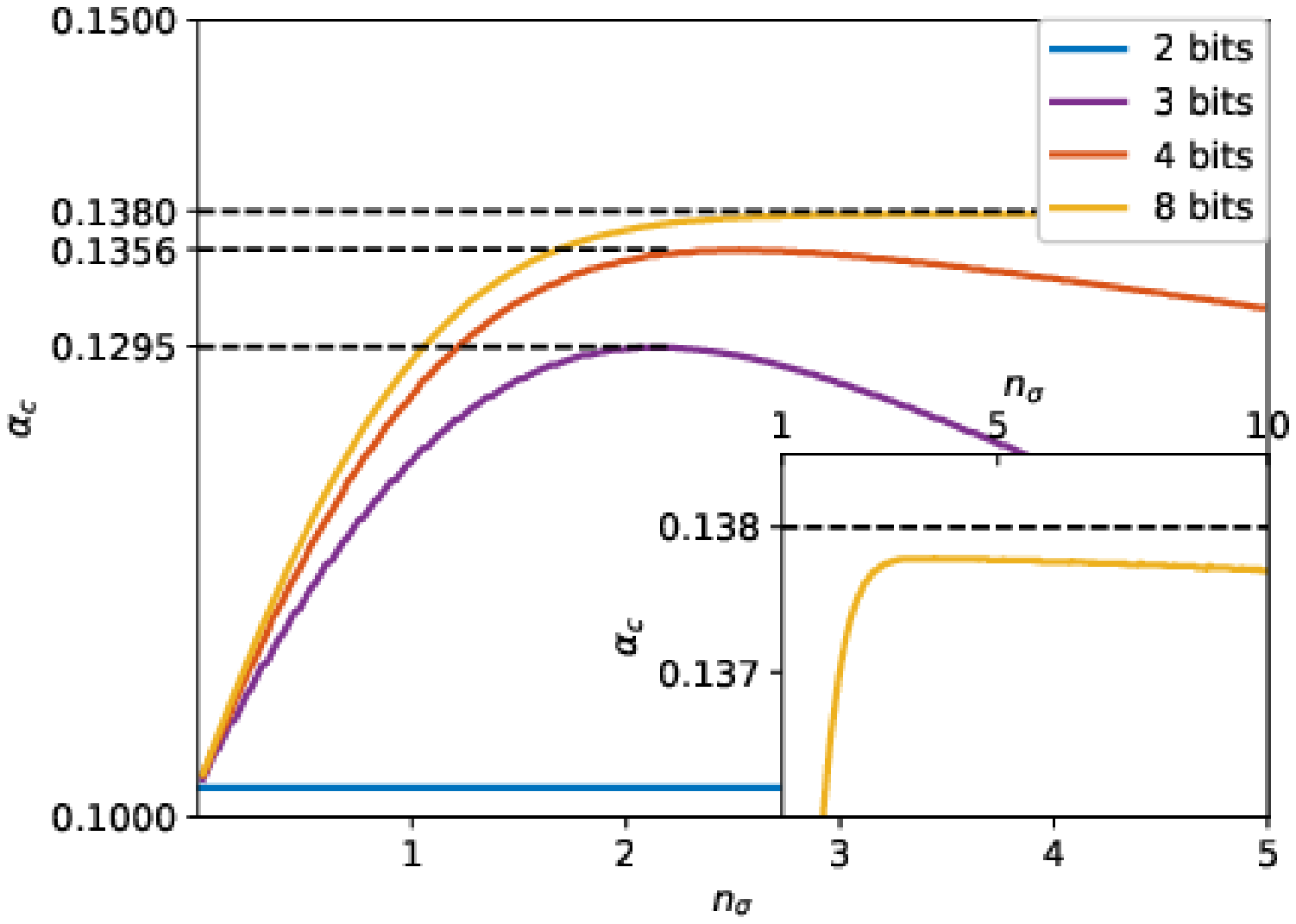}
	\label{fig:cmc_sig}
	}
	\caption{
		(a): Critical memory capacity $\alpha_c$ as a function of the number of bits
		in the limit $T \rightarrow 0$ when $n_{\sigma}=1$.
		The subfigure at the bottom right shows $\alpha_c$ as a function with a wider range of bits.
		(b): Critical memory capacity $\alpha_c$ as a function of $n_{\sigma}$ in the limit $T \rightarrow 0$.
		The subfigure at the bottom right shows $\alpha_c$ in the case of an eight-bit coupling strength as a function with a wider range of $n_{\sigma}$.
		Each figure was obtained by solving \eqref{eq:scsna_macroscopic} numerically.
	}
\end{figure*}

\subsection{Critical Memory Capacity}

From \eqref{eq:suddle_point_equations},
we can obtain the following equations
by taking the limit $T = \beta^{-1} \rightarrow 0$:
\begin{subequations}
\begin{align}
		m^1 =& {\rm erf} \left(\cfrac{m^1}{\sqrt{2\sigma^2}} \right), \\
		q =& 1, \\
		U =& \sqrt{\cfrac{2}{\pi \sigma^2}} \exp \left(- \cfrac{\left(m^1\right)^2}{2\sigma^2} \right), \\
		\sigma =& \sqrt{\left(\cfrac{1}{(1-U)^2} + \Delta^2\right)\alpha q},
\end{align}
	\label{eq:scsna_macroscopic}
\end{subequations}
where $U = J\beta(1-q)$.
These equations are identical to those obtained by SCSNA. \cite{okada1998random}
Since $q \neq 0$, the PARA phase no longer appears in the limit $T \rightarrow 0$.
These equations have a non-trivial solution with overlap $m^1\neq0$ when $\alpha<\alpha_c$.
However, when $\alpha>\alpha_c$, only the trivial solution with $m^1=0$ exists.

\figrefInit{fig:cmc_bits} shows the critical memory capacity $\alpha_c$ as a function of the number of bits
in the case of $n_{\sigma}=1$.
The critical memory capacity increase saturated after the number of bits reaches eight.
\figrefInit{fig:cmc_sig} shows the critical memory capacity $\alpha_c$ as a function of the range of the discretization function $n_{\sigma}$
in the case of two-bit, three-bit, four-bit and eight-bit coupling strengths.
In the case of a two-bit coupling strength,
$\alpha_c$ remains constant with respect to $n_{\sigma}$
since $\Delta$ is independent of the value of $n_{\sigma}$.
In the case of a three-bit coupling strength,
the critical memory capacity is maximized when $n_{\sigma}\approx2.1$,
and it decreases when $n_{\sigma}$ is more than this value.
In the case of a four-bit coupling strength,
the critical memory capacity is maximized when $n_{\sigma}\approx2.5$,
and it decreases when $n_{\sigma}$ is more than this value.
In the case of an eight-bit coupling strength,
the critical memory capacity increases until $n_{\sigma}$ approaches 3.83, 
and it decreases slowly as $n_{\sigma}$ increases.
The $\alpha_c$ obtained numerically from \eqref{eq:scsna_macroscopic} is almost equal to the value in the original Hopfield model for $n_{\sigma}\approx3.83$.

\subsection{The Almeida-Thouless Line}

To determine whether or not a replica-symmetric solution of the FM phase is stable against replica symmetry breaking (RSB), 
we calculated the Hessian matrix of the free energy.
The details were given in \appnref{sec:appendix_at_line}.
The Almeida-Thouless (AT) line \cite{de1978stability} is obtained by solving the following equations:
\begin{align}
	w^2 =& uv, 
		\label{eq:at_line} \\
	w =& 1-J^2\beta^2\Delta^2 \int Dz \cosh^{-4} J \beta (\sqrt{\alpha r + \Delta^2 q} z + m^1), \notag \\
	u =& \cfrac{1}{(1-J\beta+J\beta q)^2} + J^2 \alpha^{-1} \beta^2 \Delta^4, \notag \\
	v =& J^2\alpha\beta^2 \int Dz \cosh^{-4} J\beta (\sqrt{\alpha r + \Delta^2 q}z + m^1). \notag
\end{align}

\begin{figure}[t]
		\centering
		\includegraphics[width=0.49\textwidth]{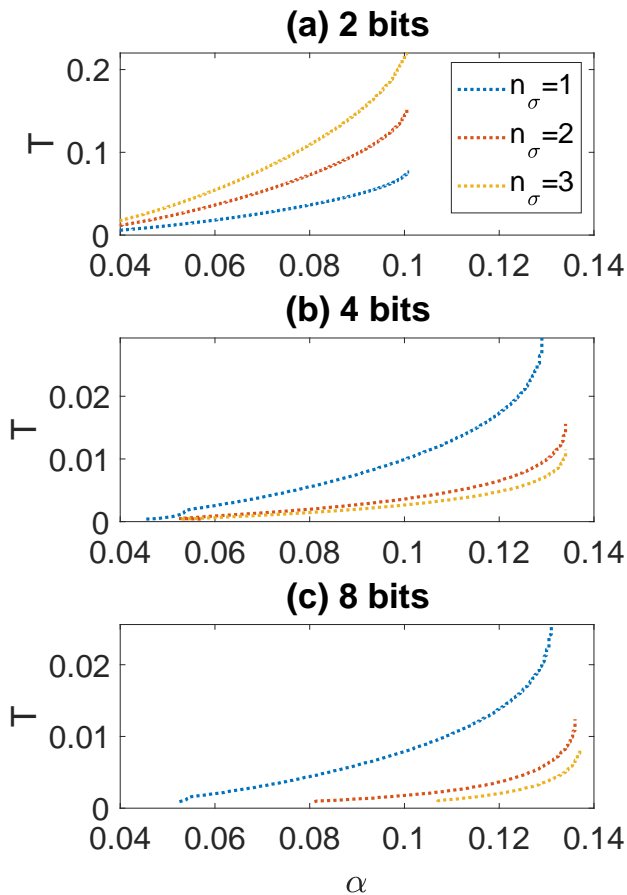}
		\caption{
			Enlarged view of the AT lines in \figref{fig:replica}.
			(a): Two-bit coupling strength.
			(b): Four-bit coupling strength.
			(c): Eight-bit coupling strength.
			Note that the scale of the vertical axis in the case of the two-bit coupling strength is different from those of the four-bit and eight-bit coupling strengths.
		}
		\label{fig:at_lines}
\end{figure}

\figrefInit{fig:at_lines} shows an enlarged view of the AT lines $T_R(\alpha)$ in \figref{fig:replica}.
These lines were obtained by numerically solving \eqref{eq:at_line}.
In the case of a two-bit coupling strength,
$T_R$ is proportional to $n_{\sigma}$,
since $\Delta$ is independent of $n_{\sigma}$.
In the cases of four-bit and eight-bit coupling strengths,
the variation in $T_R$ depending on $n_{\sigma}$ was smaller than in the two-bit case.

\section{Discussion}

We succeeded in deriving the saddle-point equations for the Hopfield model with discrete coupling by using the replica method
and used them to obtain the critical memory capacity of the model for different numbers of bits and ranges of the discretization function.
In the original Hopfield model, the critical memory capacity is 0.138. \cite{amit1985storing}
On the other hand, the critical memory capacity in the Hopfield model with clipping synapses becomes $\alpha_c = 0.1$. \cite{sompolinsky1986neural}
In \refcite{sompolinsky1986neural}, Sompolinsky showed that
the critical memory capacity $\alpha_c$ and the overlap $m^1$ increase when $\Delta$ approaches 0.
This implies that $\alpha_c$ increases by tuning the nonlinear function $f$ so that $\Delta$ becomes smaller.
It was reported that the critical memory capacity is $\alpha_c \approx 0.12$
when the three-level coupling strength taking -1, 0, or 1 was tuned such that $\Delta$ becomes the smallest. \cite{sompolinsky1987theory}
In the case of up-to-three-bit discretization,
the memory capacity was reported to be maximized by optimizing intervals non-uniformly. \cite{mimura1996}
In this study,
the memory capacity has been shown to be maximized by adjusting the range of the discretization function, $n_{\sigma}$,
depending on the number of bits in integer or fixed-point number representation.

As shown in \figref{fig:cmc_bits}, as the number of bits increases,
the critical memory capacity $\alpha_c$ monotonically increases and saturates to $0.1287$ around eight bits when $n_{\sigma} = 1$.
This result means that eight bits is sufficient to represent the coupling strength
and achieve almost the same performance as in the continuous case.
However, in the case of $n_{\sigma}=1$, the critical memory capacity does not approach $0.138$ even with numerous bits.
Thus, we also have to adjust the range of the discretization function.
As shown in \figref{fig:cmc_sig}, there is an optimal value of $n_{\sigma}$
that maximizes the critical memory capacity dependently on the number of bits.
In particular, in the case of eight bits, the critical memory capacity is maximized around $n_{\sigma} = 3$,
and it is almost the same as $0.138$.
This result shows that the model in the case of an eight-bit coupling strength with the range $n_{\sigma} = 3$
achieves almost the same performance as the original Hopfield model.
Moreover, in the case of a four-bit coupling strength with the range $n_{\sigma} = 2$,
$\alpha_c$ is degraded by about $2\%$ compared with the original Hopfield model.
In the case of a three-bit coupling strength,
the maximum value of $\alpha_c$ became $0.1295$ at around $n_{\sigma} \approx 2.1$.
This maximum memory capacity value was lower than
that of the Hopfield model with the optimal three-bit non-uniform discretization for coupling strengths
($\alpha_c = 0.135$). \cite{mimura1996}
On the other hand, in the case of a two-bit coupling strength,
$\alpha_c$ is invariant with respect to $n_{\sigma}$,
and thus, the performance can not be improved by adjusting $n_{\sigma}$ in this case.

We expect that the results obtained here give a suggestion on how many bits are needed 
to represent coupling strengths and maintain the performance of other Ising models,
because the Hopfield model shares many statistical mechanics pictures 
with other Ising models.
We surmise that the performance of other models deteriorates slightly under the four-bit condition with $n_{\sigma} = 2$,
whereas the other models under the eight-bit condition with $n_{\sigma} = 3$
achieve almost the same performance as the original ones.

\section{Conclusion}

We investigated the properties of the Hopfield model with discrete coupling.
Using the replica method,
we estimated the effect of discretization of the coupling strength on the critical memory capacity 
of the Hopfield model with discrete coupling.
As a result, 
the critical memory capacity increases as the number of bits increases.
In addition, 
we showed the relationship between the critical memory capacity and the range of the discretization function $n_{\sigma}$
and that the critical memory capacity is maximized at the optimal discretization parameter in the cases of 
three-bit,
four-bit and eight-bit coupling strengths.
In particular,
the critical memory capacity in the case of an eight-bit coupling strength and $n_{\sigma}=3$ is almost the same value as that of the original Hopfield model.
Moreover,
the critical memory capacity in the case of a four-bit coupling strength deteriorates by about 2\% in comparison with the original Hopfield model
when the range of the discretization function is optimal.
The Hopfield model shares many statistical mechanics pictures 
with other Ising models.
Thus, as discussed above,
we expect that the results obtained here give a suggestion on how many bits are needed to represent coupling strengths for maintaining the performance of other Ising models.
To achieve an efficient digital hardware implementation of Ising computing,
the number of bits representing the coupling strength should be made as small as possible
while maintaining performance as much as possible.
Our results provide reference values for designing a numerical data processor for calculating the local field.

\begin{acknowledgements}

This work is supported by the Japan Science and Technology Agency
through its ImPACT program, NTT Research Inc.,
and the National Science Foundation of the United States of America.

\end{acknowledgements}

\appendix

\section{
	\label{sec:appendix_free_energy}
	Derivation of the Free Energy
}

In this appendix, we derive the free energy using the replica method.
Using the ``replica trick,''
the average free energy per spin can be written as
\begin{align}
	f = - \lim_{n \rightarrow 0} \lim_{N \rightarrow \infty} \cfrac{\llangle [ Z^n ] \rrangle}{\beta n N}.
	\label{eq:free_energy}
\end{align}
Here, $Z$ is the partition function defined as \eqref{eq:partition_function}.
Following the recipe of the replica method,
we calculate $\llangle [ Z^n ] \rrangle$,
which is physically equivalent to the average of the partition function of $n$ replicas,
by substituting \eqref{eq:hamiltonian} and \eqref{eq:connectivity_with_static_noise}.
Substituting \seqeqref{eq:hamiltonian}{eq:connectivity_with_static_noise} into \eqref{eq:partition_function},
$\llangle [Z^n] \rrangle$ becomes
\begin{align}
	\llangle [ Z^n ] \rrangle
	=& e^{-J\beta n p / 2} \llangle \left[
		\Tr\exp \left(
			\cfrac{J\beta}{2N} \sum_{\rho=1}^n \sum_{\mu=1}^p \sum_{i,j} \xi_i^{\mu} \xi_j^{\mu} S_i^{\rho} S_j^{\rho} 
			\right. \right. \right. \right. \notag \\
			& \qquad \left. \left. \left. \left.
			+ \cfrac{\beta}{2} \sum_{\rho=1}^n \sum_{i \neq j} \eta_{ij} S_i^{\rho} S_j^{\rho}
		\right) \right] \rrangle.
\end{align}
First, we take the average over the glassy-coupling part $\eta_{ij}$.
Since $\eta_{ij}$ obeys independently and identically distributed Gaussian random variables with zero mean and variance $J^2\Delta^2/N$,
we obtain
\begin{align}
	\llangle [Z^n] \rrangle
	=& e^{-J\beta n p /2} \llangle
		\Tr \exp \left\{
			\cfrac{J\beta}{2N} \sum_{\rho=1}^n \sum_{\mu=1}^p \sum_{i,j} \xi_i^{\mu} \xi_j^{\mu} S_i^{\rho} S_j^{\rho}
			\right. \right. \right. \notag \\
			& \qquad \left. \left. \left.
			+ \cfrac{J^2\Delta^2\beta^2}{4N} \sum_{i \neq j} \left( \sum_{\rho=1}^n S_i^{\rho} S_j^{\rho} \right)
		\right\} \rrangle.
\end{align}
Next, using the standard technique in the replica method of the original Hopfield model \cite{nishimori2001statistical},
we take the quenched average over the uncondensed patterns $\{\xi_i^{\mu}\}_{\mu>1}$.
\begin{widetext}
\begin{align}
	\llangle [ Z^n ] \rrangle
	\propto& e^{-J \beta n p /2 - J^2\Delta^2\beta^2(n^2-nN)/4}
		\int \left\{ \prod_{\rho=1}^n dm_{\rho}^1 \right\}
		\int \left\{ \prod_{\rho<\sigma} dq_{\rho,\sigma} dr_{\rho,\sigma} \right\} \notag \\
		& \times \exp N \llangle
			\log \Tr \exp \beta \left\{
				J \sum_{\rho=1}^n m_{\rho}^1 \xi^1 S^{\rho}
				+ J^2 \beta \sum_{\rho<\sigma} (\alpha r_{\rho,\sigma} + \Delta^2 q_{\rho,\sigma}) S^{\rho} S^{\sigma}
			\right\}
		\rrangle_{\xi^1} \notag \\
		& \times \exp N \left(
			- \cfrac{J\beta}{2} \sum_{\rho=1}^n \left(m_{\rho}^1 \right)^2
			- J^2 \alpha\beta^2 \sum_{\rho<\sigma} r_{\rho,\sigma} q_{\rho,\sigma}
			- \cfrac{J^2\Delta^2\beta^2}{2} \sum_{\rho<\sigma} q_{\rho,\sigma}
			- \cfrac{p-1}{2} \Tr \log ((1-J\beta)\mathbb{I}_n - J \beta \mathbg{Q})
		\right), \label{eq:partition_function_power}
\end{align}
\end{widetext}
where
$\llangle\cdots\rrangle_{\xi^1}$ denotes the average over the pattern $\xi^1$,
$\mathbb{I}_n$ denotes an $n$-dimensional identity matrix,
and $\mathbg{Q}$ is a matrix whose off-diagonal elements are $q_{\rho,\sigma}$ and diagonal elements are zero.
We apply the saddle point method to the integral in \eqref{eq:partition_function_power} in the thermodynamic limit $N \rightarrow \infty$.
Accordingly, the average free energy per spin in \eqref{eq:free_energy} can be rewritten as
\begin{align}
	f=& \lim_{n \rightarrow 0} \left\{
		\cfrac{J \alpha}{2} - \cfrac{J^2\Delta^2 \beta}{4}
		+ \cfrac{\alpha}{2\beta n} \Tr \log ((1-J\beta)\mathbb{I}_n - J \beta \mathbg{Q})
		\right. \notag \\
		& \left.
		+ \cfrac{J}{2n} \sum_{\rho=1}^m \left( m^1_{\rho} \right)^2
		+ \cfrac{J^2\alpha\beta}{n} \sum_{\rho<\sigma} r_{\rho,\sigma} q_{\rho,\sigma}
		+ \cfrac{J^2\Delta^2\beta}{2n} \sum_{\rho<\sigma} q^2_{\rho,\sigma}
		\right. \notag \\
		& \left.
		- \cfrac{1}{\beta n} \llangle \log \Tr e^{\beta \ham_{\xi}} \rrangle_{\xi^1}
		\right\},
		\label{eq:general_free_energy}
\end{align}
where
\begin{align}
	\ham_{\xi}
	= J \sum_{\rho=1}^n m_{\rho}^1 \xi^1 S^{\rho}
		+ J^2 \beta \sum_{\rho<\sigma} (\alpha r_{\rho,\sigma} + \Delta^2 q_{\rho,\sigma})
		S^{\rho} S^{\sigma}.
		\label{eq:dependent_site}
\end{align}

Taking the replica symmetric ansatz,
\begin{align}
	m^1_{\rho} = m, \quad
	q_{\rho,\sigma} = q, \quad
	r_{\rho,\sigma} = r,
\end{align}
we obtain \eqref{eq:free_energy_rs}.

\section{
	\label{sec:appendix_at_line}
	Derivation of the AT Line
}

In this appendix, we derive \eqref{eq:at_line}.
The Hessian matrix of the free energy
with respect to $q_{\rho,\sigma}$ and $r_{\rho,\sigma}$
is an $n(n-1) \times n(n-1)$ matrix
around the replica-symmetric solution
having the following block structure:
\begin{align}
	C = \begin{bmatrix}
		C_{qq} & C_{qr} \\ C_{qr} & C_{rr}
	\end{bmatrix},
	\label{eq:stability_matrix}
\end{align}
where
\begin{subequations}
\begin{align}
	C_{qq}
		=& \cfrac{\del^2(nf)}{\del q_{\rho,\sigma} \del q_{\tau,\upsilon}}
		= - J^2 \alpha \beta A^{\rho\sigma,\tau\upsilon}
			- J^4 \beta^3 \Delta^4 B^{\rho\sigma,\tau\upsilon}, \\
	C_{rr}
		=& \cfrac{\del^2 (nf)}{\del r_{\rho,\sigma} \del r_{\tau,\upsilon}}
		= - J^4\alpha^2\beta^3 B^{\rho\sigma,\tau\upsilon}, \\
	C_{qr}
		=& \cfrac{\del^2 (nf)}{\del q_{\rho,\sigma} \del r_{\tau,\upsilon}}
		= J^2 \alpha \beta \delta_{\langle\rho\sigma\rangle,\langle\tau\upsilon\rangle}
			- J^4\alpha\beta^3\Delta^2 B^{\rho\sigma,\tau\upsilon}.
\end{align}
\end{subequations}
Here, $\delta_{\langle\rho\sigma\rangle,\langle\tau\upsilon\rangle}$
 takes 1 if the combination $\langle\rho\sigma\rangle$ and the combination $\langle\tau\upsilon\rangle$ are the same,
and takes 0 otherwise.

The matrices $A$ and $B$ have three different types of elements,
\begin{subequations}
\begin{align}
	A^{\rho\sigma,\rho\sigma}
		=& A_{\rho\rho}^2 + A_{\rho\sigma}^2, \\
	A^{\rho\sigma,\rho\tau}
		=& A_{\rho\rho}A_{\rho\sigma} + A_{\rho\sigma}^2, \\
	A^{\rho\sigma,\tau\upsilon}
		=& 2A_{\rho\sigma}^2,
\end{align}
	\label{eq:appendix_b_3}
\end{subequations}
where
\begin{subequations}
\begin{align}
	A_{\rho\sigma} =& \cfrac{J\beta q}{(1-J\beta+J\beta q)^2}
		\qquad (\rho\neq\sigma), \\
	A_{\rho\rho} =& \cfrac{1-J\beta+2J\beta q}{(1-J\beta+J\beta q)^2},
\end{align}
\end{subequations}
and
\begin{subequations}
\begin{align}
	B^{\rho\sigma,\rho\sigma}
		=& 1 - \langle S^{\rho} S^{\sigma} \rangle_{\ham_{\xi}}^2, \\
	B^{\rho\sigma,\rho\tau}
		=& \langle S^{\rho} S^{\sigma} \rangle_{\ham_{\xi}} - \langle S^{\rho} S^{\sigma} \rangle_{\ham_{\xi}}^2, \\
	B^{\rho\sigma,\tau\upsilon}
		=& \langle S^{\rho} S^{\sigma} S^{\tau} S^{\upsilon} \rangle_{\ham_{\xi}} - \langle S^{\rho} S^{\sigma} \rangle_{\ham_{\xi}}^2.
\end{align}
\end{subequations}
$\langle \cdots \rangle_{\ham_{\xi}}$ denotes the average by replica symmetric weight $e^{\beta\ham_{\xi}}$,
where $\ham_{\xi}$ is the Hamiltonian defined in \eqref{eq:dependent_site} under the replica symmetric ansatz.

Now let us derive the eigenvalues of the Hessian matrix \eqref{eq:stability_matrix}
in order to evaluate the stability against the following perturbation around the replica-symmetric solution.
\begin{align}
	q_{\rho,\sigma} = q + \zeta_{\rho,\sigma}, \qquad
	r_{\rho,\sigma} = r + x\zeta_{\rho,\sigma},
\end{align}
The perturbation vector $\bg{\zeta} = (\{\zeta_{\rho,\sigma}\}, \{x\zeta_{\rho,\sigma}\})$
becomes an eigenvector of the Hessian matrix \eqref{eq:stability_matrix},
which is called the replicon mode,
if $\zeta_{\rho,\sigma}$ satisfies the following condition.
\begin{align*}
	\sum_{\sigma=1}^n \zeta_{\rho,\sigma} = 0,
	\quad (\rho=1,\dots,n).
\end{align*}
Furthermore, the following conditions must be ensured for all $n$.
\begin{subequations}
\begin{align}
	\zeta_{\rho,\sigma} =& \zeta \qquad (\rho,\sigma \neq 1,2), \\
	\zeta_{1,\rho} = \zeta_{2,\rho} =& \cfrac{1}{2} (3-n)\zeta \qquad (\rho\neq1,2), \\
	\zeta_{1,2} =& \cfrac{1}{2}(2-n)(3-n)\zeta, \\
	\zeta_{\rho,\rho} =& 0
\end{align}
	\label{eq:appendix_b_7}
\end{subequations}
Let $\lambda$ be the eigenvalue corresponding to $\mathbg{\zeta}$.
The eigenvalue equation becomes
\begin{align}
	\begin{bmatrix}
		C_{qq} & C_{qr} \\ C_{qr} & C_{rr}
	\end{bmatrix} \begin{bmatrix}
		\{ \zeta_{\rho,\sigma} \} \\ \{ x\zeta_{\rho,\sigma} \}
	\end{bmatrix}
	= \lambda \begin{bmatrix}
		\{ \zeta_{\rho,\sigma} \} \\ \{ x\zeta_{\rho,\sigma} \}
	\end{bmatrix}.
\end{align}
The elementwise representation of the eigenvalue equation is given as 
\\
\begin{widetext}
\begin{align}
	&\sum_{\tau,\upsilon}\left(
		-J^2\alpha\beta A^{\rho\sigma,\tau\upsilon}
		-J^4\beta^3\Delta^4 B^{\rho\sigma,\tau\upsilon}
		-J^4\alpha\beta^3\Delta^2 B^{\rho\sigma,\tau\upsilon}x
	\right)\zeta_{\tau,\upsilon} + 2J^2\alpha\beta x\zeta
	= 2\lambda\zeta \qquad (\rho,\sigma\neq1,2), \label{eq:appendix_b_eigenvalue_equation1} \\
	&2J^2\alpha\beta\zeta + \sum_{\tau,\upsilon} \left(
		- J^4\alpha\beta^3\Delta^2B^{\rho\sigma,\tau\upsilon}
		- J^4\alpha^2\beta^3B^{\rho\sigma,\tau\upsilon}x
	\right)\zeta_{\tau,\upsilon}
	= 2\lambda x \zeta \qquad (\rho,\sigma\neq1,2).
	\label{eq:appendix_b_eigenvalue_equation2}
\end{align}
\end{widetext}
Then, substituting \eqref{eq:appendix_b_3} -- \eqref{eq:appendix_b_7} into these equations
and taking $n \rightarrow 0$,
we obtain the simultaneous equations,
\begin{align}
	&wx = \tilde{\lambda} + u,\\
	&x(\tilde{\lambda} + v) = w,
\end{align}
where
\begin{align}
	\tilde{\lambda} =& \cfrac{\lambda}{\alpha\beta}, \\
	u =& \cfrac{1}{(1-J\beta+J\beta q)^2} + J^2 \alpha^{-1} \beta^2 \Delta^4, \\
	v =& J^2\alpha\beta^2 s, \\
	w =& 1 - J^2\beta^2\Delta^2 s, \\
	s =& \int Dz \llangle \cosh^{-4} J\beta (\sqrt{\alpha r + \Delta^2 q} z + m^1\xi^1) \rrangle_{\xi^1}.
\end{align}
The eigenvalue equation becomes two-dimensional in the limit $n \rightarrow 0$;
thus, the rescaled eigenvalue $\tilde{\lambda}$ has two values, obeying
\begin{align}
	\tilde{\lambda}_{\pm}
	= \cfrac{-(u+v)\pm\sqrt{(u+v)^2+4(w^2-uv)}}{2}.
\end{align}
For any $T$, $\lambda_- <0$ holds.
On the other hand, $\lambda_+ = 0$ holds only when $w^2 = uv$.
Thus, the RSB critical point obeys the relation $w^2 = uv$,
which leads to \eqref{eq:at_line}.
Note that this is the same as the AT line of the original Hopfield model
if $J=1$ and $\Delta=0$. \cite{amit1987statistical}

\bibliography{ref}
\bibliographystyle{jpsj}

\end{document}